\documentclass[prl,aps,twocolumn,showpacs,%superscriptaddress
]{revtex4}
\usepackage{amsmath,amssymb,graphicx}
\usepackage{pxfonts}
\usepackage{graphicx}
\usepackage{color}

\usepackage{hyperref}
\hypersetup{%
	pdftitle={Excitonic Bose-polarons in electron-hole bilayers},%
	pdfauthor={E. A. Szwed et al.},%
	pdfpagemode={UseNone},%
	pdfstartview={FitH},%
	breaklinks=true,%
	citecolor=blue,%
	colorlinks=true,%
	linkcolor=blue,%
	urlcolor=blue
}

\begin{document}

\date{\today}

\title{Excitonic Bose-polarons in electron-hole bilayers}

\author{E.~A.~Szwed}
\affiliation{Department of Physics, University of California San Diego, La Jolla, CA 92093, USA}
\author{B.~Vermilyea}
\affiliation{Department of Physics, University of California San Diego, La Jolla, CA 92093, USA}
\author{D.~J.~Choksy}
\affiliation{Department of Physics, University of California San Diego, La Jolla, CA 92093, USA}
\author{Zhiwen~Zhou}
\affiliation{Department of Physics, University of California San Diego, La Jolla, CA 92093, USA}
\author{M.~M.~Fogler}
\affiliation{Department of Physics, University of California San Diego, La Jolla, CA 92093, USA}
\author{L.~V.~Butov} 
\affiliation{Department of Physics, University of California San Diego, La Jolla, CA 92093, USA}
\author{D.~K.~Efimkin}
\affiliation{School of Physics and Astronomy, Monash University, Victoria 3800, Australia}
\author{K.~W.~Baldwin}
\affiliation{Department of Electrical Engineering, Princeton University, Princeton, NJ 08544, USA}
\author{L.~N.~Pfeiffer}
\affiliation{Department of Electrical Engineering, Princeton University, Princeton, NJ 08544, USA}

\begin{abstract}
\noindent
Bose polarons are mobile impurities dressed by density fluctuations of a surrounding degenerate Bose gas. These many-body objects have been realized in ultracold atomic gasses and become a subject of intensive studies. In this work, we show that excitons in electron-hole bilayers offer new opportunities for exploring polarons in strongly interacting, highly tunable bosonic systems. We found that Bose polarons are formed by spatially direct excitons immersed in degenerate Bose gases of spatially indirect excitons (IXs). We detected both attractive and repulsive Bose polarons by measuring photoluminescence excitation spectra. We controlled the density of IX Bose gas by optical excitation and observed an enhancement of the energy splitting between attractive and repulsive Bose polarons with increasing IX density, in agreement with our theoretical calculations.

\end{abstract}
\maketitle

\begin{figure}
\begin{center}
\includegraphics[width=8.5cm]{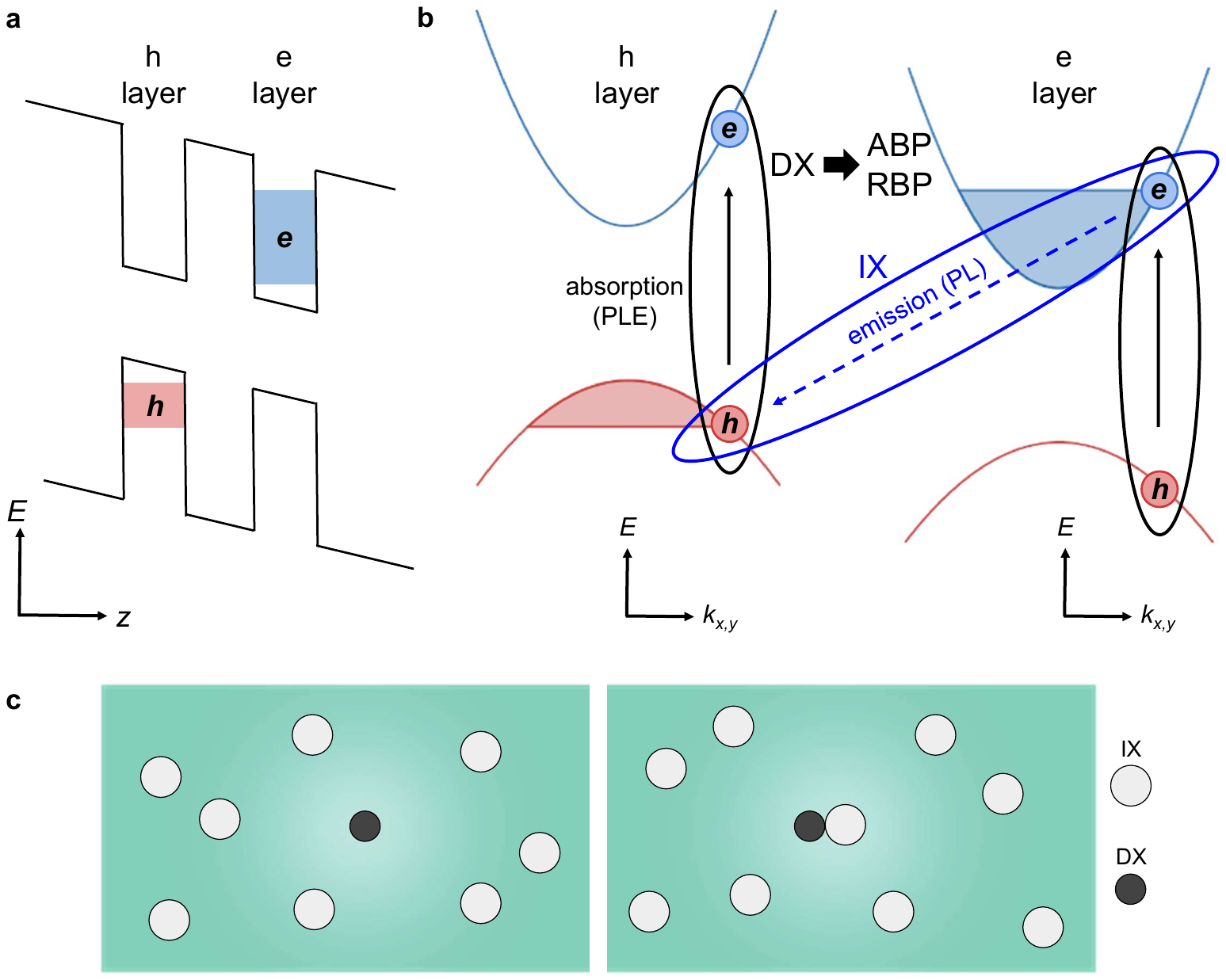}
\caption{Diagram of an electron-hole bilayer. 
(a) Electrons (e) and holes (h) are confined in separated layers. 
(b) Absorption (PLE) and emission (PL) in bilayer heterostructure. Spatially indirect excitons (IXs) and direct excitons (DXs) are shown by the ovals. 
(c) Free DXs (left) and DX-IX bound states (right) interact with surrounding IXs and form attractive and repulsive Bose polarons (ABP and RBP).
\label{fig:spectra1}}
\end{center}
\end{figure}

An impurity particle in a degenerate Bose or Fermi gas gets dressed by excitations of the surrounding medium, becoming, respectively, a Bose or Fermi polaron. Such polarons have been realized in ultracold atomic Bose~\cite{Hu2016, Jorgensen2016, Camargo2018, Yan2020, Skou2021} and Fermi~\cite{Schirotzek2009, Nascimbne2009, Kohstall2012, Koschorreck2012, Cetina2016} gases. A recent finding of excitonic Fermi polarons in degenerate electron gases in two-dimensional (2D) semiconductor heterostructures (HS) initiated intensive studies of these new quasiparticles~\cite{Sidler2017, Efimkin2017, Chang2018, Levinsen2019, Fey2020, Glazov2020, Efimkin2021, Tiene2022, Huang2023}. Degenerate Bose gases of excitons~\cite{Amelio2023} or exciton-polaritons~\cite{Tan2023} can serve as a medium for excitonic Bose polarons. In this work, we present experiments with electron-hole (e-h) bilayers in GaAs HS hosting two types of excitons: spatially indirect, or interlayer, excitons (IXs), and spatially direct, or intralayer, excitons (DXs), see Fig.~\ref{fig:spectra1}a,b. The photo-excited DXs behave as polaronic impurities in a degenerate Bose gas of IXs, Fig.~\ref{fig:spectra1}c. We observe spectroscopic signatures of two kinds of such polarons: attractive and repulsive Bose polarons (ABPs and RBPs, respectively). The ABPs are the stable low-energy polaron quasiparticles whereas RPBs are the long-lived excited states in the many-body continuum.

%%%%%%%%%%%%%%%%%%%%% FIG 1 %%%%%%%%%%%%%%%%%%%%%%%%%%%%
%%%%%%%%%%%%%%%%%%%%% FIG 1 %%%%%%%%%%%%%%%%%%%%%%%%%%%%

Semiconductor HS’s containing e-h bilayers are suited for studying ultracold neutral e-h matter (Fig.~\ref{fig:spectra1}a). The layer separation drastically increases e-h recombination lifetimes, which allows such systems to reach low-temperature quasi-equilibrium states even under optical excitation~\cite{Lozovik1976}. As summarized in Ref.~\cite{Fogler2014}, e-h bilayers can exhibit a variety of quantum phases, depending on the two key parameters: dimensionless density $n a_\mathrm{X}^2$ and dimensionless layer separation $d / a_\mathrm{X}$. Here $a_\mathrm{X} = \hbar^2 \kappa / (\mu_{\mathrm{e}\mathrm{-}\mathrm{h}} e^2)$ is the exciton Bohr radius, $\kappa$ is the dielectric constant of the semiconductor, $\mu_{\mathrm{e}\mathrm{-}\mathrm{h}} = (m_{\mathrm{e}}^{-1} + m_{\mathrm{h}}^{-1})^{-1}$ is the reduced e-h mass, and $m_\mathrm{e}$ ($m_\mathrm{h}$) is the electron (hole) effective mass. In our GaAs system $d = 19\,\mathrm{nm}$ (center-to-center separation, Fig.~\ref{fig:spectra1}a) is not much larger than $a_\mathrm{X} \approx 13\,\mathrm{nm}$, so the most experimentally relevant phases are as follows. At low densities, $n a_\mathrm{X}^2 \ll 1$, electrons and holes form a Bose-Einstein condensate (BEC) of tightly bound, hydrogen-like IXs~\cite{Keldysh1968, High2012}. At moderate densities, $n a_\mathrm{X}^2 \sim 1$, IXs become weakly bound and Cooper-pair-like, so that the BEC crosses over to a Bardeen-Cooper-Schrieffer (BCS) state~\cite{Keldysh1965, Comte1982, Choksy2023}. At high density, $n a_\mathrm{X}^2 \gg 1$, the IX binding energy is exponentially small (or perhaps, zero), so that the state of the system is best described as a correlated Fermi liquid of electrons and holes. These three regimes can be expressed as the condition on the energy scale
\begin{equation}
	E_{n} = (\pi \hbar^2 / \mu_{\mathrm{e}\mathrm{-}\mathrm{h}}) n\,,
\label{eqn:E_F_from_n}
\end{equation}
which has the meaning of the Fermi edge, i.e., the sum of the Fermi energies of noninteracting electrons and holes. This quantity is much smaller, of the order of, and larger than the IX binding energy $E_\mathrm{IX} \sim \hbar^2 / \mu_{\mathrm{e}\mathrm{-}\mathrm{h}} a_\mathrm{X}^2$ in the BEC, BCS, and Fermi-liquid regimes, respectively.

The experiments were done with GaAs HS containing two $15\,\mathrm{nm}$-thick quantum wells (QWs) separated by a $4\,\mathrm{nm}$-wide AlGaAs barrier. Electrons and holes were optically generated and their density was controlled by the laser excitation power $P_\mathrm{ex}$. Electrons and holes were driven into different QWs by an applied voltage (Fig.~\ref{fig:spectra1}a). This significantly increased e-h recombination time, to $\tau \sim 1\mu\mathrm{s}$, allowing the system to cool down and form the aforementioned quantum phases. Further details regarding the sample and the measurements are presented in Supplementary Information (SI).

Previous photoluminescence (PL) studies of this system~\cite{Choksy2023} are consistent with the BEC-BCS crossover occurring with increasing $P_\mathrm{ex}$. Here we study photoluminescence excitation (PLE) spectra. The PLE signal is a measure of optical absorption in the system. The absorption is dominated by spatially direct (intralayer) optical transitions, e.g., photoexcitation of DXs. Absorption via spatially indirect (interlayer) transitions is much weaker and is not observed. However, the DX density is always much smaller than the IX density due to a fast interlayer e-h relaxation and long IX lifetime~\cite{Choksy2021}. The IX densities in the PLE experiments were estimated to be in the range $n = 0.3\mathrm{-}11 \times 10^{10}\,\mathrm{cm}^{-2}$. These estimates are based on the measured blue shift $\delta E$ of the IX PL energy and the `capacitor' formula~\cite{Yoshioka1990} $\delta E = (4\pi e^2 d /\kappa)\, n$, which becomes increasingly more accurate as $n$ gets larger~\cite{Choksy2021, Xu2021}.

\begin{figure}
\begin{center}
\includegraphics[width=8.5cm]{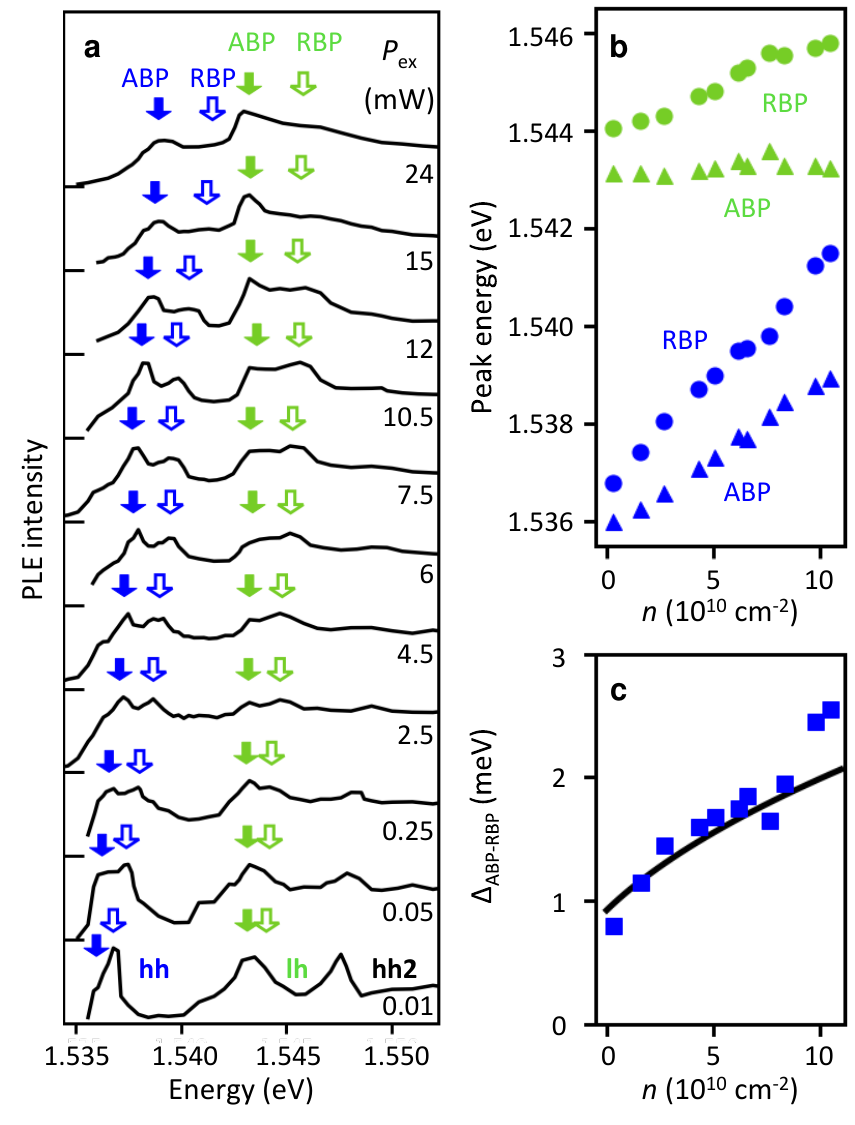}
\caption{Measured PLE spectra. (a) PLE spectra \textit{vs}. the laser excitation power $P_\mathrm{ex}$ at bath temperature $T = 2\,\mathrm{K}$. The first two peaks in the order of increasing energy correspond to ABP and RBP for DXs containing heavy holes (hh). The next two peaks correspond to ABP and RBP for DXs containing light holes (lh). The higher-energy peaks originate from higher QW subbands. 
(b) Polaron energies \textit{vs}. IX density (symbols).
(c) Measured (squares) and calculated (line) ABP-RBP energy splitting for the hh DXs.
The calculations [Eq.~\eqref{eqn:Delta}] use the estimated $E_\mathrm{XX} = (0.88 + 0.96) / 2 = 0.92\,\mathrm{meV}$ and $g_3 = 0.084 \times 10^{-10}\,\mathrm{meV}\, \mathrm{cm}^2$.
\label{fig:spectra2}}
\end{center}
\end{figure}

Our principal finding is a set of peaks in the PLE spectra, which display a pronounced $n$-dependence. As labeled in Fig.~\ref{fig:spectra2}a, we attribute the two lowest-energy peaks to ABP and RBP formed from DXs containing heavy holes (hh) and the next pair of peaks those containing light holes (lh). The photo-excited DXs behave as many-body objects dressed by excitations of the surrounding IX Bose gas. As the IX density increases, an approximately linear increase of the ABP and RBP energies (Fig.~\ref{fig:spectra2}b) as well as their splitting $\Delta_\mathrm{ABP-RBP}$ (Fig.~\ref{fig:spectra2}c) is observed. This behavior is reproduced within a theoretical model (solid line in Fig.~\ref{fig:spectra2}c) to be discussed at the end. This model treats the DX-IX interaction and phase-space filling~\cite{Schmitt-Rink1989} (PSF) effects in a unified fashion. According to this model, at low $n$ the RBP evolves into a free DX and the ABP into a DX-IX biexciton, so that the energy splitting $\Delta_\mathrm{ABP-RBP}$ should approach the DX-IX biexciton binding energy $E_\mathrm{XX}$. There are two possible biexciton types in our CQW, (h-e-h)(e) and (e-h-e)(h), where the parentheses group together particles residing in the same QW. The calculated $E_\mathrm{XX} = 0.88$ and $0.96\,\mathrm{meV}$ for, respectively, (h-e-h)(e) and (e-h-e)(h) are close to the measured $\Delta_\mathrm{ABP-RBP} = 0.8\,\mathrm{meV}$, see Fig.~\ref{fig:spectra1}c and Sec.~IV in SI.

%%%%%%%%%%%%%%%%%%%%% FIG 2 %%%%%%%%%%%%%%%%%%%%%%%%%%%%
%%%%%%%%%%%%%%%%%%%%% FIG 2 %%%%%%%%%%%%%%%%%%%%%%%%%%%%

%%%%%%%%%%%%%%%%%%%%% FIG 3 %%%%%%%%%%%%%%%%%%%%%%%%%%%%
\begin{figure}
\begin{center}
\includegraphics[width=8.5cm]{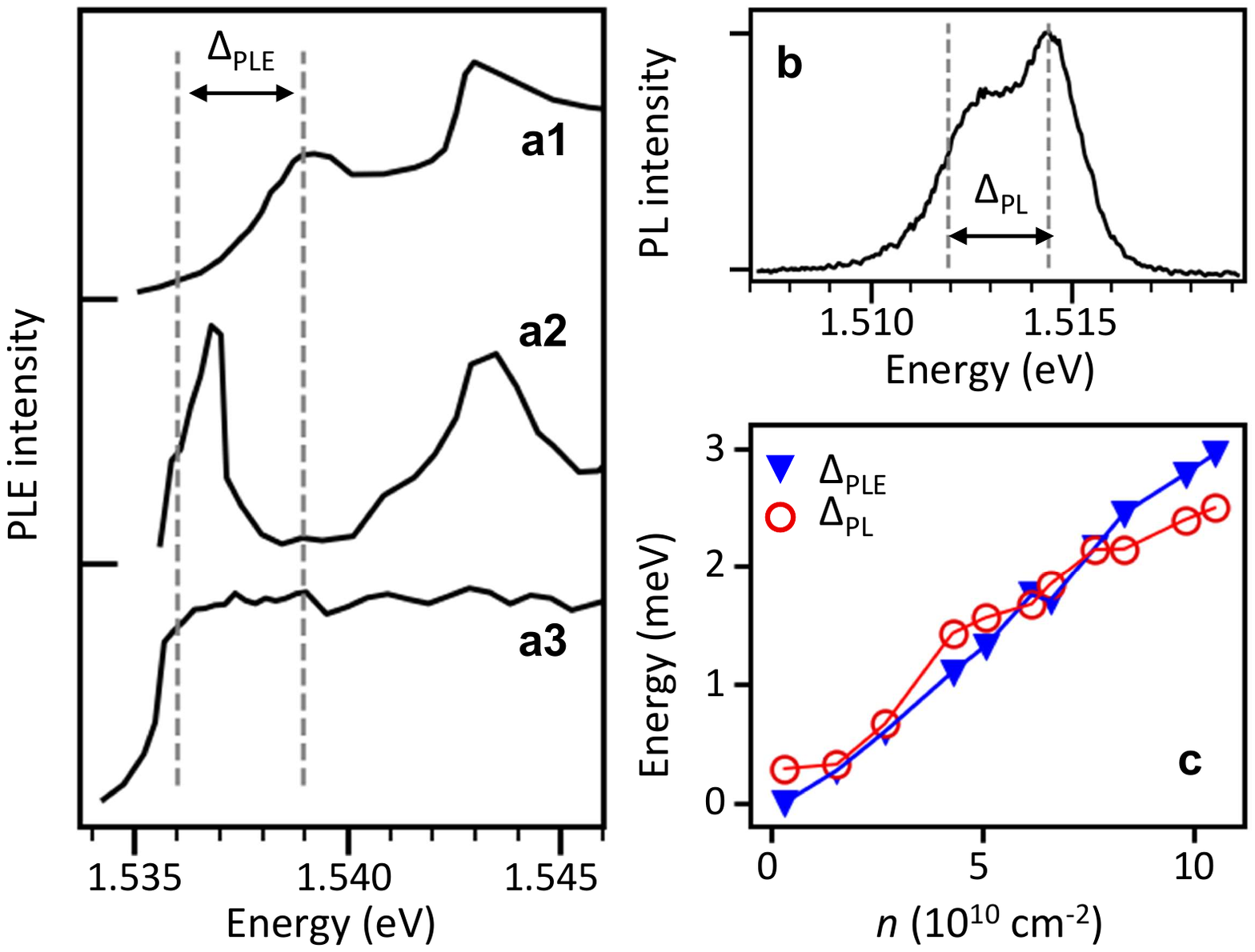}
\caption{PLE energy shift and PL energy width. 
(a) PLE spectra at
$T = 2$~K, $P_\mathrm{ex} = 24$~mW (line a1);
$T = 2$~K, $P_\mathrm{ex} = 0.01$~mW (line a2);
$T = 20$~K, $P_\mathrm{ex} = 24$~mW (line a3).  
At high temperatures (line a3), the excitonic peaks have vanished and the spectrum has become step-like, corresponding to the density of states of a 2D e-h plasma. $\Delta_\mathrm{PLE}$ denotes the shift of the hh-ABP line with increasing density.
(b) PL spectrum at $T = 2$~K, $P_\mathrm{ex} = 24\,\mathrm{mW}$, estimated e-h density $n = 1.1 \times 10^{11}$~cm$^{-2}$. The PL enhancement at the Fermi edge comes from Cooper-pair-like IXs~\cite{Choksy2023}. $\Delta_\mathrm{PL}$ is the PL energy of Cooper-pair like IX at the Fermi energy relative to the low-energy PL.
(c) $\Delta_\mathrm{PL}$ and $\Delta_\mathrm{PLE}$ \textit{vs}. IX density. $T = 2$~K.
The observed relation $\Delta_\mathrm{PLE} \approx \Delta_\mathrm{PL}$ indicates that the ABP energy shift is close to $E_{n}$. 
\label{fig:spectra3}
}
\end{center}
\end{figure}
%%%%%%%%%%%%%%%%%%%%% FIG 3 %%%%%%%%%%%%%%%%%%%%%%%%%%%%

In Fig.~\ref{fig:spectra3} we compare two energy scales: the energy shift $\Delta_\mathrm{PLE}$ of the lowest-energy ABP peak in the PLE spectra (Fig.~\ref{fig:spectra2}a) and the spectral width $\Delta_\mathrm{PL}$ of the PL line (Fig.~\ref{fig:spectra3}b). We measure $\Delta_\mathrm{PLE}$ with respect to the lowest-density ABP position $1.536\,\mathrm{eV}$. At the higher densities, a shoulder is observed at this energy in the PLE spectra marking the onset of absorption. This shoulder is more pronounced at $P_\mathrm{ex} = 2.5\mathrm{-}7.5\,\mathrm{mW}$ (Fig.~\ref{fig:spectra2}a) and its position is nearly density independent. In turn, we define $\Delta_\mathrm{PL}$ as the distance from the shoulder on the low-energy side at half-maximum intensity and a peak on the high-energy side of the PL line (Fig.~\ref{fig:spectra3}b). The peak appears at high enough densities and is a signature of Cooper-pair-like excitons (Fig.~\ref{fig:spectra3}b) in the BCS phase~\cite{Choksy2023}. Apart from that peak, the PL has the box-like shape expected for cold e-h plasma, so $\Delta_\mathrm{PL}$ is approximately equal to the Fermi edge $E_{n}$. (We used this consideration to verify the accuracy of the capacitor formula for $n$.) Figure~\ref{fig:spectra3}c indicates that $\Delta_\mathrm{PLE}$ and $\Delta_\mathrm{PL}$ nearly coincide in most of the studied density range. This implies that $\Delta_\mathrm{PLE} \approx E_{n}$, i.e., the energy shift of the ABP peak is of the order of Fermi edge energy. As alluded to above, this energy shift comes from a combination of Coulomb interaction and PSF effects, see more at the end.

With increasing temperature, the quantum-degenerate Bose gas of IXs evolves into classical IX gas at low densities and e-h plasma at high densities, respectively. The ABP and RBP lines vanish and the PLE spectrum becomes step-like (Fig.~\ref{fig:spectra3}, line a3). The step-like PLE and the box-like PL have the same origin: the density of states being flat in 2D. At high temperatures, the absorption edge in the dense e-h system drops in comparison to low temperatures (Fig.~\ref{fig:spectra3}, lines a1 and a3), as expected for the reduction of the PSF in a classical e-h gas~\cite{Schmitt-Rink1989}. 

At the highest e-h densities in our experiments, the Fermi edge energies $E_{n}$ are above the IX binding energy $E_\mathrm{IX} = 3\,\mathrm{meV}$ (SI, Sec.~IV) and, therefore, IXs are in the high-density regime. Still $E_{n}$ are below the calculated DX binding energy $E_\mathrm{DX} = 8$--$9\,\mathrm{meV}$ (SI, Sec.~IV), and so the polaron approach remains sensible.

We now discuss higher-energy peaks in the PLE spectra (Fig.~\ref{fig:spectra2}a). We attribute the third and fourth peaks (in the order of increasing energy) to lh DXs. Indeed, at the lowest IX density, the separation $\sim 7\,\mathrm{meV}$ between this pair of peaks and the first pair of peaks agrees with the calculated lh-hh splitting~\cite{Bastard1986}, see SI for details. At finite density, the lh DXs become Bose polarons and their ABP-RBP energy splitting increases with $n$ similarly to the case of hh DXs (Fig.~\ref{fig:spectra2}a,b). However, for lh lines the energy increase with $n$ is slower than for the hh lines (Fig.~\ref{fig:spectra2}a,b). This suggests that the PSF and Coulomb interaction effects are weaker for lh polarons compared to hh polarons when the hh IX states are occupied.

The PLE peaks seen at even higher energies originate from higher-order QW subbands. For instance, we assign the peak at $1.548\,\mathrm{eV}$ (Fig.~\ref{fig:spectra2}a) to hh2, the heavy holes in the second subband. We do so because the splitting of approximately $12\,\mathrm{meV}$ of these peaks from the hh DX peaks (Fig.~\ref{fig:spectra2}a) matches the calculated splitting between the hh and hh2 subbands in $15\,\mathrm{nm}$ QWs~\cite{Bastard1986} (see also SI). The hh2 peaks disappear at relatively low densities (Fig.~\ref{fig:spectra2}a) that limits their analysis.

Now we discuss our theoretical model of excitonic polarons. In this model, the ABP corresponds to the DX-IX biexciton dressed by excitations of the IX Bose gas while the RBP corresponds to a dressed free DX. Our numerical study of four-body systems (SI, Sec.~IV) reveals that biexcitons are stable only in the symmetric channel, e.g., when the total spin of the two identical particles in the same layer [electrons for (e-h-e)(h) and holes for (e)(h-e-h)] is zero. Accordingly, at low IX densities $n$, we can treat DXs and IXs as composite bosons interacting via a short-range effective potential, which is binding in the symmetric and non-binding in the antisymmetric channel. At the mean-field level, the $n$-dependence of the polaron energies is linear:
$E_\mathrm{RBP}^{(0)}(n) = n g_1$,
$E_\mathrm{ABP}^{(0)}(n) = n g_2 - E_\mathrm{XX}$.
These energies are referenced to the bare DX energy. Parameter $g_1$ can be estimated within the Hartree-Fock approximation~\cite{Schmitt-Rink1989, Ciuti1998, Schindler2008}, which includes the Coulomb interaction and PSF effects to the lowest order and yields
\begin{equation}
	 n g_1 = 0.45 E_{n}
\label{eqn:g_1}
\end{equation}
for the hh case (SI, Sec.~VII). We also expect $g_2 \approx g_1$, at least, away from ultralow density regime. The idea is that since the DX-IX biexciton is weakly bound, being `tethered' to one of IXs does not affect the interaction of the DX with other IXs too much.

An improvement over the naive mean-field theory is the $T$-matrix approximation where the polariton energy is derived from the equation $E = \mathrm{Re}\,\Sigma(E)$, with $\Sigma(E) = n T(E)$ being the DX self-energy~\cite{Schmitt-Rink1989, Haug1984, Schindler2008}. The essential physics of the problem can be modelled by the $T$-matrix of the Fano-Feshbach form
\begin{equation}
	T(E) = g_1 + g_3 \frac{E_\mathrm{XX}}{E - E_\mathrm{ABP}^{(0)}}.
	\label{eqn:T-matrix}
\end{equation}
We take $g_i$'s to be constant, thus neglecting any energy or momentum dependence of DX-IX scattering. We also assume them to be real, so we neglect the imaginary part of $T(E)$ and the corresponding broadening of the spectral lines as well. Parameter $g_3$ characterizes the width of the DX-IX bound-state resonance. Parameter $g_1$, already introduced above, represents non-resonant DX-IX scattering. For $g_1 = g_2$ case this model predicts 
\begin{align}
	E_\mathrm{ABP,RBP} &= n g_1 - \frac12 E_\mathrm{XX}
	\pm \frac12 \Delta_\mathrm{ABP-RBP} \,,
\label{eqn:mean-field1}	\\
	\Delta_\mathrm{ABP-RBP} &= \left(E_\mathrm{XX}^2
	+ 4 n g_3 E_\mathrm{XX}
	\right)^{1/2} .
\label{eqn:Delta}
\end{align}
%%%
Hence, the $n$-dependence of the energy splitting $\Delta_\mathrm{ABP-RBP}$ is controlled by $g_3$. For a quick estimate, we choose $g_3 = \hbar^2 (\pi / 2) (1 / m + 1 / m_\mathrm{DX})$, which gives the correct residue of the pole in $T(E)$ in the limit $n \to 0$ (SI, Sec.~V). With $m_\mathrm{DX} = m \equiv m_\mathrm{e} +  m_\mathrm{h} = 0.285 m_0$ for the hh exciton mass we find $g_3 = 0.084 \times 10^{-10}\,\mathrm{meV}\, \mathrm{cm}^2$. The corresponding $\Delta_\mathrm{ABP-RBP}(n)$ is shown by the line in Fig.~\ref{fig:spectra2}c. It agrees reasonably well with the experiment. Next, we find $g_1 = 0.21 \times 10^{-10}\,\mathrm{meV}\, \mathrm{cm}^2$ from Eq.~\eqref{eqn:g_1}. In comparison, by fitting Eq.~\eqref{eqn:mean-field1} to the two lower data sets in Fig.~\ref{fig:spectra2}b, i.e., hh ABP and RBP, we obtained $g_1 = 0.34 \times 10^{-10}\,\mathrm{meV}\, \mathrm{cm}^2$ (SI, Sec.~VII). This level of agreement is rather good for such crude estimates.

As mentioned in the introduction, the Mott transition (or crossover) from the excitonic to the correlated e-h Fermi liquid regime is expected to occur in our system at high $n$. The corresponding transition from Bose to Fermi polarons may result, which can be an interesting topic for future work.

In summary, we presented spectroscopic evidence for excitonic Bose polarons in electron-hole bilayers. These polarons are many-body objects formed around spatially direct excitons in a degenerate Bose gas of spatially indirect excitons. The energy splitting between attractive and repulsive branches of the Bose polarons grows with the indirect exciton density. We interpreted this behavior within a theoretical model employing the estimated biexciton binding energy and exciton interaction parameters.

\section{Acknowledgments}
\acknowledgments

We thank E. Demler, J. Levinsen, and M. Parish for discussions. The PLE studies were supported by DOE Award DE-FG02-07ER46449, the PL studies by NSF Grant 1905478, and the heterostructure growth by Gordon and Betty Moore Foundation Grant GBMF9615 and NSF grant DMR 2011750.

\section{References}
%\begin{references}

\end{document}

\date{\today}

\title{Supporting Information for
``Excitonic Bose-polarons in electron-hole bilayers''}

\author{E.~A.~Szwed}
\affiliation{Department of Physics, University of California San Diego, La Jolla, CA 92093, USA}
\author{B.~Vermilyea}
\affiliation{Department of Physics, University of California San Diego, La Jolla, CA 92093, USA}
\author{D.~J.~Choksy}
\affiliation{Department of Physics, University of California San Diego, La Jolla, CA 92093, USA}
\author{Zhiwen~Zhou}
\affiliation{Department of Physics, University of California San Diego, La Jolla, CA 92093, USA}
\author{M.~M.~Fogler}
\affiliation{Department of Physics, University of California San Diego, La Jolla, CA 92093, USA}
\author{L.~V.~Butov}
\affiliation{Department of Physics, University of California San Diego, La Jolla, CA 92093, USA}
\author{D.~K.~Efimkin}
\affiliation{School of Physics and Astronomy, Monash University, Victoria 3800, Australia}
\author{K.~W.~Baldwin}
\affiliation{Department of Electrical Engineering, Princeton University, Princeton, NJ 08544, USA}
\author{L.~N.~Pfeiffer}
\affiliation{Department of Electrical Engineering, Princeton University, Princeton, NJ 08544, USA}

\begin{abstract}
\noindent
\end{abstract}

\maketitle
\renewcommand*{\thefigure}{S\arabic{figure}}
\renewcommand*{\theequation}{S\arabic{equation}}
\renewcommand*{\thetable}{S\,\Roman{table}}

\section{Sample}
\label{sec:sample}

The sample used in this study is a coupled quantum well (CQW) heterostructure grown by molecular beam epitaxy.
The CQW consists of two $15$-$\mathrm{nm}$-wide GaAs QWs separated by a $4$-$\mathrm{nm}$-thick Al$_{0.33}$Ga$_{0.67}$As barrier. An $n^+$ GaAs layer with $n_\mathrm{Si} \sim 10^{18}\,\mathrm{cm}^{-3}$ serves as a uniform bottom gate.
The CQW is positioned $100\,\mathrm{nm}$ above the $n^+$ GaAs layer within the undoped $1$-$\mu\mathrm{m}$-thick Al$_{0.33}$Ga$_{0.67}$As layer. The CQW is located
much closer to the bottom gate 
%is 
to minimize the effect of fringing electric fields 
%that arise 
in excitonic devices with patterned top gates~\cite{Hammack2006}. The top semi-transparent gate is fabricated by applying $2\,\mathrm{nm}$ of Ti and $7\,\mathrm{nm}$ of Pt on a $7.5\,\mathrm{nm}$-thick GaAs cap layer.
Applied gate voltage $V_\mathrm{g} = -2.5\,\mathrm{V}$ creates an electric field in the direction normal to the quantum wells.
The corresponding band diagram of the CQW is shown in Fig.~1a of the main text. 
The applied voltage drives optically generated electrons (e) and holes (h) to the opposite quantum wells. This process is fast, so that the densities of minority particles (e's in the h-layer and h's in the e-layer) are orders of magnitude smaller than the densities of majority particles (e's in the e-layer and h's in the h-layer).

\begin{figure}
	\begin{center}
		\includegraphics[width=5cm]{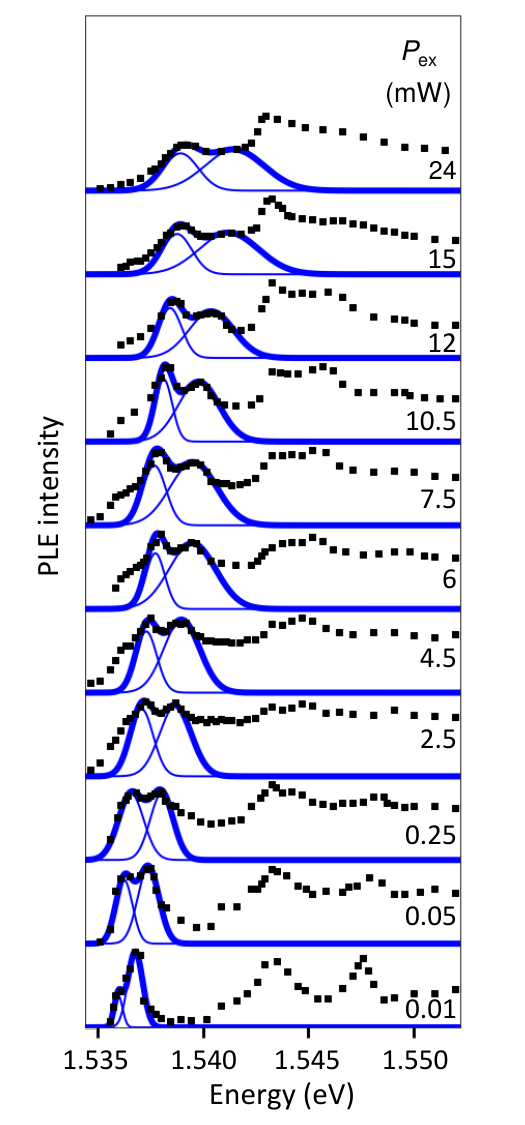}
		\caption{Gaussian fits to the ABP and RBP peaks in the measured PLE spectra (points, same as Fig.~2a of the main text). The individual fits are shown by the thin lines and their sum by the thick line. The peak energies found from the fits are presented in Fig.~2b of the main text.
			\label{fig:spectra2}
		}
	\end{center}
\end{figure}

\section{Optical measurements details}

The PLE spectra (Fig.~\ref{fig:spectra2}) probe spatially direct optical absorption within each QW. A spatially indirect absorption is much weaker and is not observed in PLE. The PL and PLE were measured $\sim 50\,\mu\mathrm{m}$ away from the laser excitation spot and $\sim 300\,\mathrm{ns}$ after the excitation pulse, where a cold and dense e-h system of temperature close to the lattice temperature was formed~\cite{Choksy2023}. 
To facilitate comparison with prior PL measurements~\cite{Choksy2023}, we use similar optical excitation and detection protocol, as follows.
The e-h system is generated by a Ti:\,Sapphire laser. An acousto-optic modulator is used for making laser pulses ($800\,\mathrm{ns}$ on, $400\,\mathrm{ns}$ off).
A laser excitation spot with a mesa-shaped intensity profile and
diameter $\sim 100\,\mu\mathrm{m}$ is created using an axicon.
The signal is detected within a $50\,\mathrm{ns}$ window,
which is much shorter than the IX lifetime, so that the signal variation during the measurement remains negligible~\cite{Choksy2023}. 
The exciton density in the detection region is close to the density in the excitation spot because the separation is shorter than the IX propagation length and the time delay is shorter than the IX lifetime~\cite{Choksy2023}.
The IX PL spectra are measured using a spectrometer with resolution $0.2\,\mathrm{meV}$ and a liquid-nitrogen-cooled CCD coupled to a PicoStar HR TauTec time-gated intensifier.
The experiments are performed in a variable-temperature $^4$He cryostat.

%\begin{figure}
%\begin{center}
%\includegraphics[width=3.5cm]{figS1.pdf}
%\caption{Schematic of electron-hole (e-h) complexes in a CQW.
%Indirect excitons (IXs), direct excitons (DXs), and two types of DX-IX biexcitons are shown by the ovals.
%%			
%% Indicate layer thickness ($15\,\mathrm{nm}$) and center-to-center distance
%% $d = 19\,\mathrm{nm}$, redraw everything to scale.
%\label{fig:spectra1}
%}
%\end{center}
%\end{figure}

\section{PLE spectra}

We used Gaussian fits for rough estimates of the ABP and RBP peak energies (Fig.~\ref{fig:spectra2}). The actual ABP and RBP PLE lineshapes are complicated.
In particular, their low-energy sides appearing
near $1.536\,\mathrm{eV}$ have a non-Gaussian shoulder-like form (Fig.~\ref{fig:spectra2}).
The analysis of the lh-ABP and lh-RBP lines is also challenging because
they appear on a background of optical transitions between free heavy holes and electrons, see the main text. Nevertheless, the variation of all the observed polaron energies with density are sufficiently strong and systematic (Fig.~2 of the main text).
We note that the fit accuracy is lower for the highest $P_{\rm ex}$, in particular, due to the RBP peak broadening.

\begin{figure}
\begin{center}
\includegraphics[width=7.5cm]{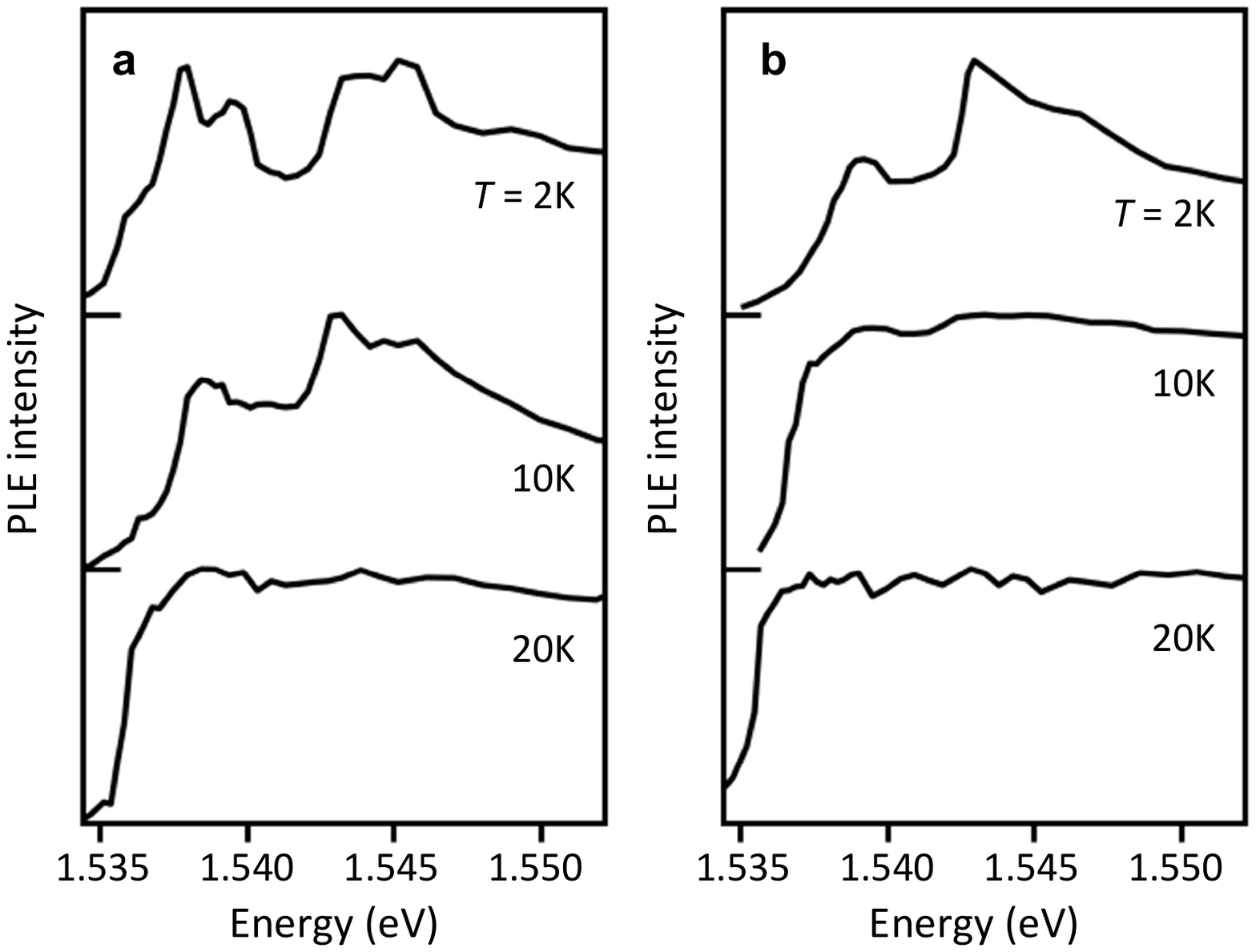}
\caption{PLE spectra at $P_{\rm ex} = 7.5$ (a) and 24 (b) mW at $T = 2$, 10, and 20~K. At high temperatures, the excitonic peaks vanish and the PLE spectrum becomes step-like as 2D density of states.
\label{fig:spectra3}
}
\end{center}
\end{figure}

With increasing temperature, the ABP and RBP lines vanish and the PLE spectrum becomes step-like, 
reflecting the functional form of the e-h joint density of states in 2D (Fig.~\ref{fig:spectra3}). At high temperatures, the absorption edge in the dense e-h system decreases compared to the low-temperature data (Fig.~\ref{fig:spectra3}), as discussed in the main text.

\section{Exciton binding energies}
\label{sec:dispersion}

Few-body e-h bound states that can form in the CQW are
%  depicted in Fig.~\ref{fig:spectra1}.
listed in Table~\ref{tbl:energies}, together with their calculated binding energies.
They include indirect excitons (IXs), direct excitons (DXs), and DX-IX biexcitons.
The details of the calculations are presented below.

%%%%%%%%%%%%%%%%%%%%% TABLE 1 %%%%%%%%%%%%%%%%%%%%%%%%%%%%
\begin{table}[bth]
	\begin{ruledtabular}
		\begin{tabular}{lllcc}
			Complex & QW 1 & QW 2 & \textit{h} $ = $ hh & \textit{h} $ = $ lh
			\\
			\hline 
			IX & e & hh & $2.99$ & 
			\\[3pt]
			DX & e-\textit{h} &  & $8.24$ & $9.44$
			\\[3pt]
			DX-IX & e-\textit{h}-e & hh & $0.96$ & $1.26$
			\\[3pt]
			DX-IX & e & \textit{h}-e-hh & $0.88$ & $0.73$
		\end{tabular}
	\end{ruledtabular}
	\caption{Calculated binding energies of various e-h complexes, in units of $\mathrm{meV}$, for zero gate voltage $V_\mathrm{g} = 0$; 'hh' and 'lh' stand for the heavy hole and light hole, respectively.
		\label{tbl:energies}
	}
\end{table}
%%%%%%%%%%%%%%%%%%%%% TABLE 1 %%%%%%%%%%%%%%%%%%%%%%%%%%%%

\begin{figure}
	\begin{center}
		\includegraphics[width=6cm]{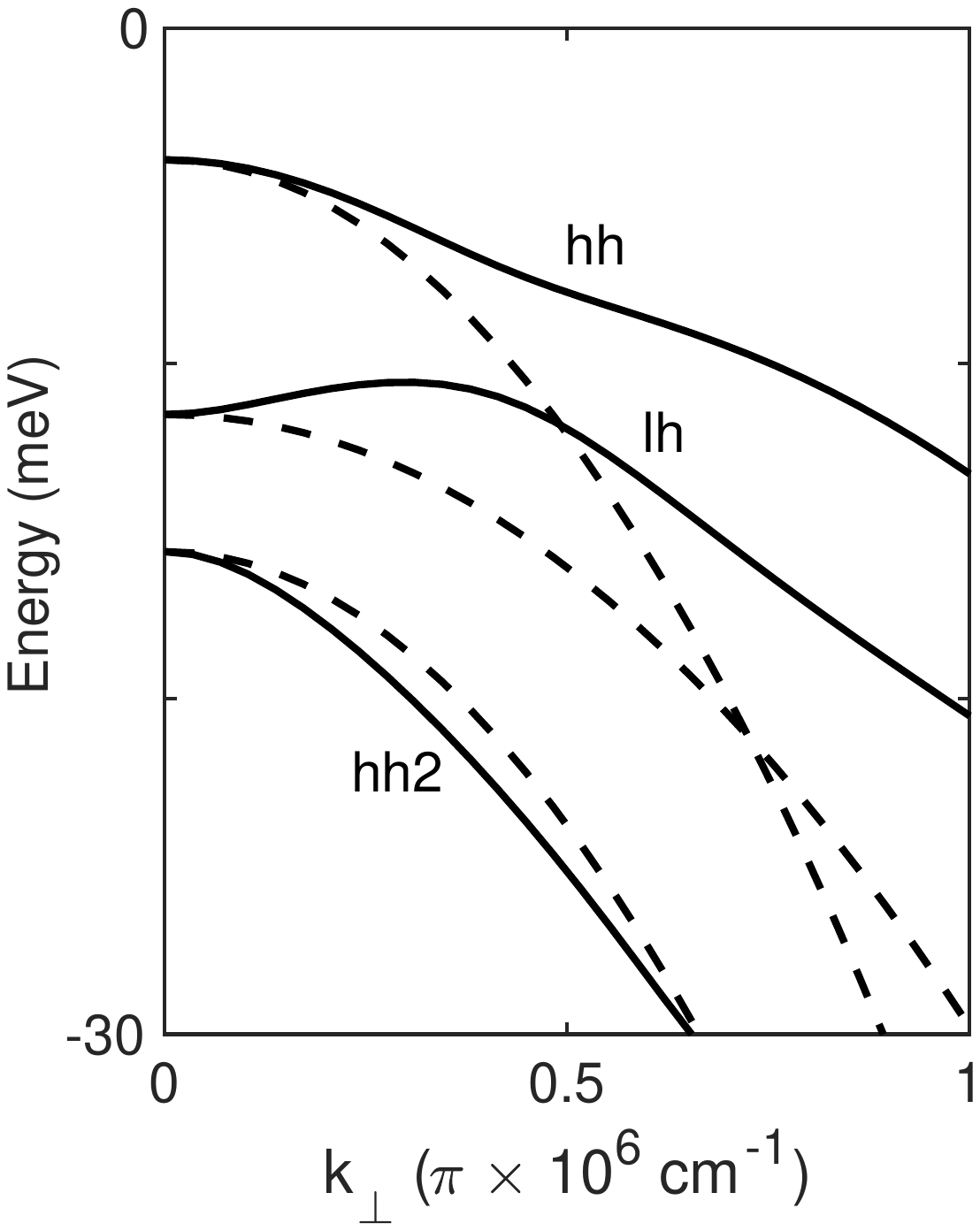}
		\caption{In-plane dispersion $\varepsilon_\mathrm{h}(\mathbf{k}_\bot)$ for the first three hole subbands in units of $\mathrm{meV}$.
			The dashed lines are the dispersions obtained by neglecting
			the coupling between heavy and light holes (the off-diagonal
			terms of $H_h$).
			\label{fig:bands}
		}
	\end{center}
\end{figure}

% In order to calculate the exciton and biexciton binding energies,
% we first solved
The first step of the calculation is to solve
for the single-particle states of the QWs.
The electron states were determined from the Hamiltonian
\begin{equation}
	H_\mathrm{e} = \frac{1}{2 m_\mathrm{e}} {\mathbf{P}}^2 + U_\mathrm{e}(z)\,,
	\label{eqn:H_e}
\end{equation}
where $z$ is the coordinate perpendicular to the QW plane,
${\mathbf{P}} = (\mathbf{p}, -i\hbar \partial_z)$ is the momentum operator,
$\mathbf{p} = \hbar \mathbf{k}_\bot$ is the in-plane momentum,
and $m_\mathrm{e} = 0.0665 m_0$ is the effective electron mass in GaAs.
The hole states were determined from the Hamiltonian~\cite{Bastard1986, Vasko1998}
\begin{equation}
\begin{split}
H_\mathrm{h} &= -\frac{1}{2 m_0} \sum_{ij} P_i D_{ij} P_j + U_\mathrm{h}(z),
\\
\hat D_{ij} &= \left(\frac12 \gamma_1 + \frac54 \gamma_2\right)
\delta_{ij} - \gamma_2 {J_i} {J_j},
\end{split}
\label{eqn:H_h}
\end{equation}
where $\gamma_1$ and $\gamma_2$ are the Luttinger parameters,
$m_0$ is the free electron mass,
and ${\mathbf{J}} = (J_x, J_y, J_z)$ is the spin-$3/2$ angular momentum operator.
The confining potentials $U_\mathrm{e}(z)$ and $U_\mathrm{h}(z)$ were chosen in the form
\begin{equation}
U_\mathrm{e, h}(z) =
\begin{cases}
	0 & \text{inside QW},\\
	\mathcal{U}_\mathrm{e, h} & \text{outside QW}.
\end{cases}
\end{equation}
For all the parameter values in our calculations we used those given in Ref.~\onlinecite{Sivalertporn2012}.
We numerically diagonalized the Hamiltonians in Eqs.~\eqref{eqn:H_e} and \eqref{eqn:H_h} 
and obtained the energy levels and wavefunctions $\varphi_i(z)$ where $i =
\mathrm{e(h)} n$ for e(h) states of 
QW $n = 1, 2$.
The energy-momentum dispersions of the first three hole subbands are plotted
in Fig.~\ref{fig:bands}.
To facilitate comparison with published results~\cite{Bastard1986, Vasko1998}
(that appears to be good)
the momentum on the horizontal axis is expressed in units of $\pi \times 10^{6}\,\mathrm{cm}^{-1}$.

%%%%%%%%%%%%%%%%%%%%%%%%%%%%%%%%%%%%%%%%%%%%%%%%%%%%%%%%%%%%%
%%%%%%%%%%%%%%%%%%%%%%%%%%%%%%%%%%%%%%%%%%%%%%%%%%%%%%%%%%%%%

Next, to define the effective mass $m_\mathrm{h}$ of the heavy hole (hh),
we fitted its dispersion to a parabola
over a range of momenta $0 < k_\bot < a_\mathrm{X}^{-1}$, where
$a_\mathrm{X} = (\kappa \hbar^2/ e^2)(m_\mathrm{e}^{-1} + m_\mathrm{h}^{-1})$ is again the exciton Bohr radius.
We found $m_\mathrm{h} = 0.217 m_0 = 3.26 m_\mathrm{e}$,
so that $a_\mathrm{X} = 12.7\,\mathrm{nm}$. Note that $a_\mathrm{X}^{-1}$ is about $0.25$ in the
momentum units used in Fig.~\ref{fig:bands}.
The light hole (lh) dispersion is non-monotonic.
For simplicity, we decided to neglect this dispersion altogether,
i.e., to treat the lh mass as infinite.

To compute the binding energies of interest we approximated the momentum-space Coulomb interaction potential between particles of charge $e_i$ and $e_j$ by
\begin{equation}
\widetilde{V}_{ij}(\mathbf{k}_\bot) = \frac{2\pi e_i e_j}{\kappa {k}_\bot}
\int dz dz'|\varphi_i(z)|^2\, |\varphi_j(z')|^2\, e^{-{k}_\bot |z-z'|}\,,
\label{eqn:V}
\end{equation}
which we further simplified as follows.
For particles in the same layer, we used~\cite{Vasko1998}
\begin{subequations} \label{eqn:V_approx}
\begin{gather}
\widetilde{V}_{ij}(\mathbf{k}_\bot) =
\frac{2\pi e_i e_j}{\kappa {k}_\bot}\,
\frac{1}{1 + {k}_\bot \rho_{ij}}\,,
\label{eqn:V_q_approx}\\
V_{ij}(\mathbf{r}) =
    \frac{\pi}{2 \rho_{ij}}\,
    \frac{e_i e_j}{\kappa}\,
	\left[
	\mathbf{H}_0\left(\frac{r}{\rho_{ij}}\right) - Y_0\left(\frac{r}{\rho_{ij}}\right)
	\right],
\label{eqn:V_r_approx}
\end{gather}
\end{subequations}
where the effective well widths $\rho_{\mathrm{e}n,\mathrm{e}n} = 4.5\,\mathrm{nm}$, $\rho_{\mathrm{h}n,\mathrm{h}n} = 3.81\,\mathrm{nm}$, 
and $\rho_{\mathrm{e}n,\mathrm{h}n} = 4.17\,\mathrm{nm}$ (all for hh),
were determined by numerically evaluating the integrals in Eq.~\eqref{eqn:V}
and fitting the result to Eq.~\eqref{eqn:V_q_approx} at $0 < {k}_\bot < a_\mathrm{X}^{-1}$.
Equation~\eqref{eqn:V_r_approx} is known as the Rytova-Keldysh potential.
This function approaches the Coulomb potential $e_i e_j / \kappa r$ at $r \gg \rho_{ij}$ and
diverges logarithmically $(e_i e_j / \kappa \rho_{ij}) \ln(\rho_{ij} / r)$ at $r \ll \rho_{ij}$; 
$\mathbf{H}_0(z)$ and $Y_0(z)$ are the Struve and Neumann functions, respectively. 

For particles in opposite layers, we used
$\rho_{i j} = 0$, i.e., the Coulomb law:
\begin{subequations} \label{eqn:V_Coulomb}
\begin{gather}
\widetilde{V}_{ij}(\mathbf{k}_\bot) = 2\pi \frac{e_i e_j}{\kappa {k}_\bot}\, e^{-{k}_\bot d},
\label{eqn:V_ij_q_Coulomb}\\
V_{ij}(\mathbf{r}) = \frac{e_i e_j}{\kappa}\, \frac{1}{\sqrt{r^2 + d^2}}\,,
\label{eqn:V_ij_r_Coulomb}
\end{gather}
\end{subequations}
where $d = 19\,\mathrm{nm}$ is the center-to-center layer distance.
These interlayer and intralayer potentials are plotted in Fig.~\ref{fig:V_IX}.
We neglected intersubband mixing
because the energy separation between the subbands is relatively large,
$5$--$7\,\mathrm{meV}$, see Fig.~\ref{fig:bands}.

We computed the DX and IX binding energies $E_\mathrm{X}$ and ground-state
wavefunctions $\phi_\mathrm{X}(\mathbf{k}_\bot)$ 
by numerically solving the Wannier equation,
%%
%\begin{equation}
\begin{align}
	[\varepsilon_\mathrm{e}(\mathbf{k}_\bot) + \varepsilon_\mathrm{h}(\mathbf{k}_\bot)]\phi_\mathrm{X}(\mathbf{k}_\bot)
	&+ \Omega^{-1} \sum_{\mathbf{k}'_\bot} \widetilde{V}_{\mathrm{e}k,\mathrm{h}n}(\mathbf{k}_\bot - \mathbf{k}'_\bot) \phi_\mathrm{X}(\mathbf{k}'_\bot)
	\notag\\
	&= -E_\mathrm{X} \phi_\mathrm{X}(\mathbf{k}_\bot)
	\label{eqn:H_X}
\end{align}
%\end{equation}
%%
following Ref.~\onlinecite{Chao1991}.
Here $\mathrm{X} \in \{\mathrm{DX}, \mathrm{IX}\}$ is the exciton type,
$\varepsilon_\mathrm{e,h}(\mathbf{k}_\bot) = \hbar^2 \mathbf{k}^2_\bot / 2 m_\mathrm{e,h}$
are the e(h) dispersions,
and $\Omega$ is the area of the system.

Finally, we calculated the biexciton binding energies using the stochastic variational method (SVM),
a highly accurate numerical technique for solving few-body quantum mechanics problems~\cite{Varga1995}.
To this end we adopted the SVM code previously developed~\cite{Meyertholen2008}
for zero-thickness 2D layers ($\rho_{ij} \equiv 0$) and modified it to
work with the interaction potential of Eq.~\eqref{eqn:V_approx}.
We also used the SVM solver to verify the exciton binding energies $E_\mathrm{X}$
computed by the diagonalization method and found them to be in excellent agreement.
Table~\ref{tbl:energies} summarizes
the results for all the binding energies we calculated.

%%%%%%%%%%%%%%%%%%%%%%%%%%%%%%%%%%%%%%%%%%%%%%%%%%%%%%%%%%%%%%%%%%%%%%%%%%%%%%%%%%%%%%
%%%%%%%%%%%%%%%%%%%%%%%%%%%%%%%%%%%%%%%%%%%%%%%%%%%%%%%%%%%%%%%%%%%%%%%%%%%%%%%%%%%%%%
%%%%%%%%%%%%%%%%%%%%%%%%%%%%%%%%%%%%%%%%%%%%%%%%%%%%%%%%%%%%%%%%%%%%%%%%%%%%%%%%%%%%%%
\section{Exciton-exciton interaction}
\label{sec:interaction}

\subsection{IX-IX interaction}
\label{sub:IX-IX}

%%%%%%%%%%%%%%%%%%%%%%%%%%%%%%%%%%%%%%%%%%%%%%%%%%%%%%%%%%%%%
\begin{figure}
	\begin{center}
		\includegraphics[width=6cm]{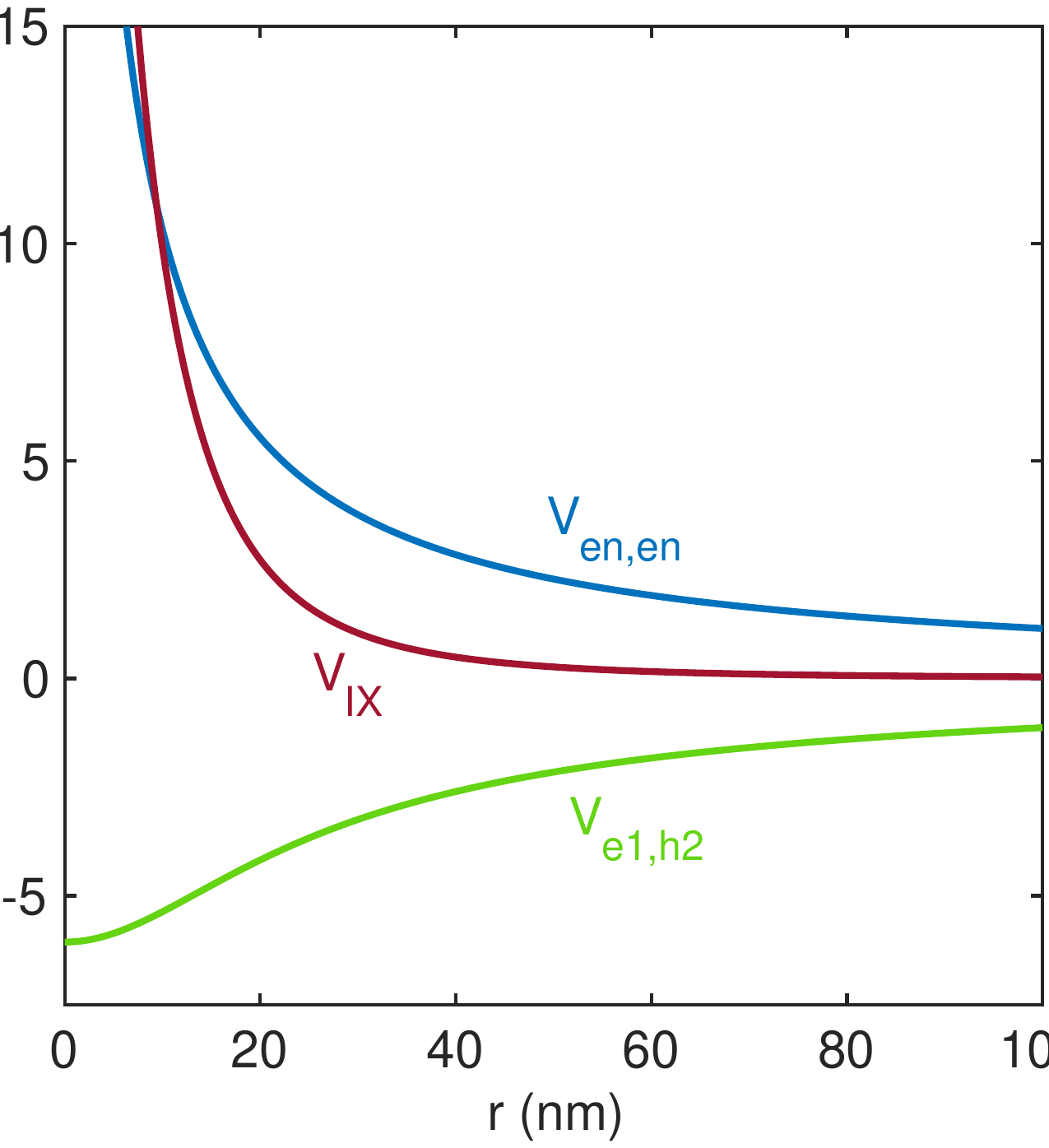}
		\caption{Model interaction potentials: intralayer potential $V_{\mathrm{e}n,\mathrm{e}n} = V_{\mathrm{h}n,\mathrm{h}n}$ [Eq.~\eqref{eqn:V_r_approx}],  interlayer potential $V_{\mathrm{e}1,\mathrm{h}2}$ [Eq.~\eqref{eqn:V_ij_r_Coulomb}], and the IX-IX potential $V_\mathrm{IX}$ [Eq.~\eqref{eqn:V_IX}].
			\label{fig:V_IX}
		}
	\end{center}
\end{figure}
%%%%%%%%%%%%%%%%%%%%%%%%%%%%%%%%%%%%%%%%%%%%%%%%%%%%%%%%%%%%%

Theoretical investigations of Bose polarons have been stimulated primarily by experiments with cold atoms.
Transferring these methods to excitons must be done with caution because of important differences between two classes of systems.
Atoms reach quantum degeneracy at very low temperatures
in the n$\mathrm{K}$ or $\mu\mathrm{K}$-range.
Since IXs have much smaller mass, their degeneracy temperature
$T_\mathrm{deg} = 2\pi T_0$ with $T_0 \equiv \hbar^2 n / m$~\cite{Fogler2014}
is many orders of magnitude higher, e.g.,
$T_\mathrm{deg} = 20\,\mathrm{K}$ for $m = m_\mathrm{e} + m_\mathrm{h} = 0.285 m_0$ 
and $n = 10^{11}\,\mathrm{cm}^{-2}$.

Atoms typically form dilute, weakly nonideal BECs (Bose-Einstein condensates) 
for which the details of the interatomic interaction potential are unimportant.
Instead, the interactions are parametrized by the $s$-wave scattering amplitude,
which is proportional to the on-shell two-body $T$-matrix.
In 2D, this $T$-matrix has a universal low-energy form
\begin{equation}
	T(E) \simeq \frac{4\pi \hbar^2}{m} \left[\ln\left(-\frac{\hbar^2}{m a^2}\, \frac{1}{E}\right)\right]^{-1},
	\quad E \to 0,
	\label{eqn:T_2D}
\end{equation}
where $a$ is referred to as the scattering length.
The $T$-matrix enters in equations for
many key quantities of the system.
For example, the chemical potential $\zeta$ of the BEC is given by
\begin{equation}
	\zeta = T n
	\label{eqn:zeta_from_n}
\end{equation}
to the leading order in $n a^2 \ll 1$.
In this equation, $T$ needs to be evaluated at energy
$E \sim -\zeta$~\cite{Schick1971, Popov1972},
so that Eq.~\eqref{eqn:zeta_from_n} is a self-consistent equation for $\zeta$ as a function of $n$.
The solution can be presented in the form
\begin{equation}
	\tilde{g} \equiv \frac{\zeta}{T_\mathrm{0}}
	= \frac{m}{\hbar^2}\, \frac{\zeta}{n}
	\simeq \frac{4\pi}{\ln \left({1} / {n a^2}\right)}\,.
	\label{eqn:diluteness}
\end{equation}
The dimensionless parameter $\tilde{g}$ is a measure of BEC nonideality.
For example, it determines the interaction-induced condensate depletion via $\tilde{g} / (2\pi)$~\cite{Popov1972}.
These formulas apply if
$\tilde{g} \lesssim 1$~\cite{Astrakharchik2007, Astrakharchik2009}, which translates to
the condition on the boson density $n a^2 \lesssim 10^{-6}$.
Despite the small numerical factor on the right-hand side
of this inequality,
it is not uncommon to have it fulfilled for cold atoms.
In contrast, such densities are unrealistically low for IXs in GaAs heterostructures.
As a result, scattering length $a$ is not useful for describing these excitonic systems.
Their properties crucially depend on details of the IX-IX interaction
and they are typically strongly coupled, $\tilde{g} \gg 1$.

One common model for the interaction potential of two IXs is
\begin{equation}
V_{\mathrm{IX}}(r) = V_\mathrm{ee}(r) + V_\mathrm{hh}(r) + 2 V_\mathrm{eh}(r)\,,
\label{eqn:V_IX}
\end{equation}
where $r$ is the distance between the centers of mass of the IXs.
As one can see from Fig.~\ref{fig:V_IX},
potential $V_{\mathrm{IX}}(r)$ has a strong repulsive core and rapidly decreasing tails.
Equation~\eqref{eqn:V_IX} is essentially classical, e.g.,
it neglects fermionic and bosonic exchange of IXs~\cite{Haug1984}
at distances $r \lesssim a_\mathrm{X}$.
However, due to the strong IX-IX repulsion~\cite{Schindler2008, Meyertholen2008},
excitons tend to avoid each other and these exchange effects should be small
at densities $n \ll a_\mathrm{X}^{-2} \sim 6 \times 10^{11}\,\mathrm{cm}^{-2}$
studied in our experiments.
% in our sample where $d / a_\mathrm{X} \approx 1.5$.
% (In contrast, fermionic exchange between DX and IX is important, see Sec.~\ref{sub:DX-IX} below.)

At $r \gg d$, the IX-IX potential approaches $V_{\mathrm{IX}}(r) \simeq {e^2 d^2} / {\kappa r^3}$.
The corresponding $s$-wave scattering length $a$ is given by~\cite{Astrakharchik2009} $a = e^{2\gamma_E} d^2 / A_{X}$,
where $\gamma_E = 0.577$ is the Euler constant and
$A_\mathrm{X} = \hbar^2 \kappa / m e^2 = 2.3\,\mathrm{nm}$.
For $d = 19\,\mathrm{nm}$, we find $a = 500\,\mathrm{nm}$, so that in our experiments $n a^2 \gg 1$.
In this regime Eq.~\eqref{eqn:diluteness} fails and is replaced by
\begin{subequations} \label{eqn:g_0_IX}
\begin{numcases}{\tilde{g} \equiv \frac{m}{\hbar^2}\, \frac{\zeta}{n} =}
C_g \sqrt{n a^2}\,, &  $n \ll 1 / d^2$,
	\label{eqn:g_0_dipole}\\
4\pi e^{-2\gamma_E} \frac{a}{d}\,, & $n \gg 1 / d^2$,
\label{eqn:g_0_capacitor}
\end{numcases}
\end{subequations}
which is specific to the interaction law~\eqref{eqn:V_IX}.
The numerical constant $C_g \sim 5$ in Eq.~\eqref{eqn:g_0_dipole} can be estimated from Ref.~\cite{Xu2021} and work cited therein.
Note that Eq.~\eqref{eqn:g_0_capacitor} is the same as the `capacitor formula' introduced in the main text.
From these equations, we find $\zeta \sim 30$--$300\,\mathrm{K}$, $T_0 \sim 0.3$--$3\,\mathrm{K}$, and $\tilde{g} \sim 40$--$80$ in our experiments, indicating that
IXs form a strongly correlated Bose gas rather than a weakly nonideal BEC.
The large value of $\tilde{g}$ is not a cause for concern; it simply shows that the $s$-wave scattering length $a$ is not a meaningful control parameter for such dense many-body systems.

\subsection{DX-IX interaction}
\label{sub:DX-IX}

The interaction between impurities and host bosons in cold atom gases and in excitonic 
systems also has some qualitative differences.
In the context of cold atoms it is common
to describe this interaction using another parameter of dimension of length --- the size of the impurity-host dimer.
If this length is much larger than the scattering length $a$ of the host bosons,
the impurity can attract many host particles.
As a result, the ABP becomes a multi-particle cluster with energy much lower than the dimer energy~\cite{Pena_Ardila2020}.
A related effect is formation of multimers (trimers, quadrimers, \textit{etc}.) in a few-body bosonic systems~\cite{Guijarro2020}.
In our case, the size of the DX-IX bound state, defined by the relation
\begin{equation}
	a_\mathrm{XX} = \frac{\hbar}{\left(2\mu E_\mathrm{XX}\right)^{1/2}} \approx 15\,\mathrm{nm},
	\label{eqn:a_DX-IX}
\end{equation}
is much smaller than the IX-IX scattering length $a \approx 500\,\mathrm{nm}$.
[Here $\mu = (m_\mathrm{DX}^{-1} + m^{-1})^{-1}$ is the reduced mass of DX and IX,
$m$ is the IX mass, and $m_\mathrm{DX}$ is the DX mass. We used $E_\mathrm{XX} = 1.11\,\mathrm{meV}$, which is the average of the hh and lh values in Table~\ref{tbl:energies}.]
This means that the IX-IX repulsion is strong compared to the DX-IX attraction.
Therefore no multimers or multi-exciton clusters 
can appear and the excitonic ABP is essentially a dimer.

As mentioned in the main text, the DX-IX bound states, e.g., (e-h-e)(h) biexcitons, which Eq.~\eqref{eqn:a_DX-IX} refers to,
are stable only when the spins on the two e's form a singlet.
The spin dependence of the interaction of the excitons
comes from the symmetries of their orbital wavefunctions.
It indicates that exchange plays an important role in the DX-IX
interaction unlike the case of the IX-IX interaction discussed in Sec.~\ref{sub:IX-IX} above.

The exchange effects can be analyzed as follows.
Taking the (e-h-e)(h) complex as an example, we note that in GaAs each of the four particles
involved can exist in two spin states, $s_z = \pm 1/2$ for the e's and
$J_z = \pm 3/2$ for the h's, yielding $2^4 = 16$ combinations total.
In this Hilbert space we can select a basis of spin wavefunctions  that are either even or odd with respect to interchange of e's or h's.
The corresponding orbital wavefunctions must have the opposite parity and therefore different scattering amplitudes.
Following Ref.~\cite{Schindler2008}, we can describe the DX-IX interaction using four different $T$-matrices
$T_{v}^{u}$, where  $u$ and $v$ refer to e and h, respectively, $u, v \in \{\mathrm{s}, \mathrm{a}\}$ and $\mathrm{s}$($\mathrm{a}$) indicates symmetric (antisymmetric) orbital wavefunction.
The $u = v = \mathrm{s}$ channel is a singlet.
The spin degeneracy triples if $u$ or $v$ is switched from $\mathrm{s}$ to $\mathrm{a}$,
so that the original $16$-fold degeneracy is split into four channels of spin degeneracy $1$, $3$, $3$, and $9$.
In the present case, the problem is actually simpler because we can neglect exchange between particles residing in different QWs,
e.g., the h-exchange in the (e-h-e)(h) DX-IX complex.
Thus, we can disregard the spin of the two h's.
We need to consider only the four e-spin states that split into
an antisymmetric triplet, described by a $T$-matrix $T_\mathrm{a}^\mathrm{a} = T_\mathrm{s}^\mathrm{a} \equiv T^\mathrm{a}$ and a symmetric singlet, characterized by another $T$-matrix $T_\mathrm{a}^\mathrm{s} = T_\mathrm{s}^\mathrm{s} \equiv T^\mathrm{s}$.

Some properties of these $T$-matrices are known from general principles.
The triplet channel is non-binding, the singlet channel supports bound state(s).
Therefore, $T^\mathrm{a}(E)$ is analytic at all negative energies $E < 0$ whereas $T^\mathrm{s}(E)$ has a pole at $E = -E_\mathrm{XX}$.
In the asymptotic low-energy limit $E \to 0$,
both $T^\mathrm{a}$ and $T^\mathrm{s}$ have the universal form
[cf.~Eq.~\eqref{eqn:T_2D}]
\begin{gather}
	T^u(E) = \frac{\widetilde{V}^u}{1 - L(E) \widetilde{V}^u},
	\label{eqn:T^u}\\
	L(E) = \frac{1}{\Omega} \sum_{|\mathbf{k}| < \Lambda}
	\frac{1}{E - \varepsilon_\mathbf{k} - \varepsilon_{\mathrm{DX}, \mathbf{k}}}
	%\notag\\
	\simeq -\frac{\mu}{2\pi \hbar^2} 
	 \ln \left(-\frac{\hbar^2 \Lambda^2}{2\mu E}\right).
%	 \ln \left(-\frac{\hbar^2 \Lambda^2}{2\mu}\, \frac{1}{E}\right).
	\label{eqn:L}
\end{gather}
This expression represents the sum of all ladder diagrams
for two particles --- an IX with dispersion
$\varepsilon_\mathbf{k} = \hbar^2 \mathbf{k}^2 / 2m$ and a DX with dispersion $\varepsilon_{\mathrm{DX}, \mathbf{k}} = \hbar^2 \mathbf{k}^2 / 2m_\mathrm{DX}$ --- interacting via a short-range effective potential $V^{u}(r)$ such that
$\widetilde{V}^\mathrm{a} > 0$ and $\widetilde{V}^\mathrm{s} < 0$.
Parameter $\Lambda \sim a_\mathrm{X}^{-1}$ is the high-momentum cutoff. 
If the binding energy $E_\mathrm{XX}$ belongs to the range of validity of Eq.~\eqref{eqn:T^u},
then $\widetilde{V}^\mathrm{s}$ can be deduced from the condition that
$T^\mathrm{s}(E)$ has a pole at $E = -E_\mathrm{XX}$:
\begin{equation}
	\widetilde{V}^\mathrm{s} = -\frac{\hbar^2}{\mu}\,
	\frac{\pi}{\ln \left(\Lambda a_\mathrm{XX}\right)}\,,
	\label{eqn:g^s_from_a_XX}
\end{equation}
which entails
\begin{equation}
\begin{split}
	T^\mathrm{s}(E) &= \frac{2\pi \hbar^2}{\mu} \left[\ln\left(-\frac{E_\mathrm{XX}}{E}\right)\right]^{-1}
	\\
	&\simeq \frac{2\pi \hbar^2}{\mu} \,
	\frac{E_\mathrm{XX}}{E + E_\mathrm{XX}}\,,
	\quad E \to -E_\mathrm{XX}\,.
	\label{eqn:T^s}
\end{split}
\end{equation}
Accurate calculation of $T^\mathrm{a}$ and $T^\mathrm{s}$ at arbitrary energies and momenta requires solving the four-body scattering problem numerically, which goes beyond the scope of the present work.
(Currently, our numerical codes can only solve for the bound states,
see Sec.~\ref{sec:dispersion}.)
However, we can estimate $T^\mathrm{a}$ and $T^\mathrm{s}$
by combining Eqs.~\eqref{eqn:T^u}, \eqref{eqn:L}
with the Hartree-Fock approximation
for $\widetilde{V}^\mathrm{a(s)} = \widetilde{V}_d \pm \widetilde{V}_x$~\cite{Haug1984, Schmitt-Rink1989, Schindler2008}.
Due to the exciton charge neutrality,
the Hartree (or direct) term $\widetilde{V}_d$ is negligible compared to
the Fock (or e-exchange) term $\widetilde{V}_x$, so that
\begin{equation}
	T^\mathrm{a(s)}(E) \approx \pm \frac{\widetilde{V}_x}{1 \mp L(E) \widetilde{V}_x}\,.
	\label{eqn:T^u_HF}
\end{equation}
The equation for the Fock term is
\begin{equation}
	\begin{split}
		\widetilde{V}_x &= -\int \frac{d^2k}{(2\pi)^2} \int \frac{d^2k'}{(2\pi)^2}
		W(\mathbf{k}, \mathbf{k}')\,,
		\\
		W(\mathbf{k}, \mathbf{k}')
		&= \widetilde{V}_\mathrm{e1, e1}(\mathbf{k} - \mathbf{k}')
		\Phi(\mathbf{k}, \mathbf{k}'; \mathbf{k}, \mathbf{k}')
		\\
		&+ \widetilde{V}_\mathrm{e1, h2}(\mathbf{k} - \mathbf{k}')
		\Phi(\mathbf{k}, \mathbf{k}; \mathbf{k}, \mathbf{k}')
		\\
		&+ \widetilde{V}_\mathrm{h1, e1}(\mathbf{k} - \mathbf{k}')
		\Phi(\mathbf{k}, \mathbf{k}; \mathbf{k}', \mathbf{k})
		\\
		&+ \widetilde{V}_\mathrm{h1, h2}(\mathbf{k} - \mathbf{k}')
		\Phi(\mathbf{k}, \mathbf{k}'; \mathbf{k}', \mathbf{k})
		\,,
		\label{eqn:V_x}
	\end{split}
\end{equation}
where
\begin{equation}
	\Phi(\mathbf{k}, \mathbf{q}; \mathbf{k}'\!, \mathbf{q}') \equiv \phi_\mathrm{DX}^*(\mathbf{k}) \phi_\mathrm{IX}^*(\mathbf{q}) \phi_\mathrm{DX}(\mathbf{k}') \phi_\mathrm{IX}(\mathbf{q}')\,.
	\label{eqn:Phi}
\end{equation}
For comparison with previous work, we can write
\begin{equation}
	\widetilde{V}_x = C_x \frac{\hbar^2}{2 \mu_{\mathrm{e-h}}}\,,
	\quad
	\frac{1}{\mu_{\mathrm{e-h}}} \equiv \frac{1}{m_\mathrm{e}} + \frac{1}{m_\mathrm{h}}.
	\label{eqn:V_x_est}
\end{equation}
Using $\phi_\mathrm{DX}(\mathbf{k})$, $\phi_\mathrm{IX}(\mathbf{k})$
found as described in Sec.~\ref{sec:dispersion}, we obtained
the numerical coefficients $C_x = 3.81$ for hh and $3.24$ for lh.
Interestingly,
they are only slightly larger than the analytical result $C_x = 4\pi - (315 \pi^3 / 1024) = 3.03$ for the DX-DX interaction in a zero-thickness QW~\cite{Schmitt-Rink1989}.
In physical units, we find
\begin{equation}
\widetilde{V}_x = 0.28 \times 10^{-10}\,\mathrm{meV}\,\mathrm{cm}^{2}
\label{eqn:V_x_hh}
\end{equation}
for (e-h-e)(h) with h $ = $ hh.

At this point we can compare the Hartree-Fock estimate
$\widetilde{V}^\mathrm{s} \approx -\widetilde{V}_x$ with
Eq.~\eqref{eqn:g^s_from_a_XX}.
In fact, we can get them to agree perfectly by fixing the numerical factor in the momentum cutoff parameter, making the `large logarithm' in Eq.~\eqref{eqn:g^s_from_a_XX} equal to $\ln\left(\Lambda a_\mathrm{XX}\right) = 0.59$, which corresponds to $\Lambda = 1.5 / a_\mathrm{X}$.
With this adjustment, Eq.~\eqref{eqn:T^s}
for $\widetilde{T}^\mathrm{s}$ reproduces the accurate value of the binding energy $E_\mathrm{XX} = 0.96\,\mathrm{meV}$ in Table~\ref{tbl:energies}.
It may now be tempting to use Eq.~\eqref{eqn:T^u_HF} for $\widetilde{T}^\mathrm{a}$
with the same $\Lambda$.
However, doing so would generate a spurious pole in $\widetilde{T}^\mathrm{a}(E)$ at a relatively small (by absolute value) energy
\begin{equation}
	E = -\left(\frac{\hbar^2 \Lambda^2}{2\mu}\right)^2 \frac{1}{E_\mathrm{XX}}
	 \approx -10\,\mathrm{meV}\,.
	\label{eqn:E_spurious}
\end{equation}
We believe it is a sign of going beyond the range of validity of the approximation.
Therefore, it may be better to revert to the lowest-order perturbation theory formula
\begin{equation}
	T^\mathrm{a}(E) = \widetilde{V}_x = \mathrm{const}\,.
	\label{eqn:T^a_HF}
\end{equation}
We take Eqs.~\eqref{eqn:T^s}, \eqref{eqn:V_x_hh}, and \eqref{eqn:T^a_HF}
for two-body DX-IX scattering as the basis for the further analysis of the many-body Bose polaron problem in Sec.~\ref{sec:g-ology}.

%%%%%%%%%%%%%%%%%%%%%%%%%%%%%%%%%%%%%%%%%%%%%%%%%%%%%%%%%%%%%%%%%%%%%%%%%%%%%%%%%%%%%%
%%%%%%%%%%%%%%%%%%%%%%%%%%%%%%%%%%%%%%%%%%%%%%%%%%%%%%%%%%%%%%%%%%%%%%%%%%%%%%%%%%%%%%
%%%%%%%%%%%%%%%%%%%%%%%%%%%%%%%%%%%%%%%%%%%%%%%%%%%%%%%%%%%%%%%%%%%%%%%%%%%%%%%%%%%%%%
\section{Bose polarons in weakly interacting 2D systems}
\label{sec:BEC}

There have been numerous theoretical studies of Bose polarons in all physical dimensions:
3D, 2D, and 1D.
Some examples of methods developed to tackle the 2D case with short-range interactions
include the Fr\"ohlich polaron model, which was treated by the Feynman variational method~\cite{Casteels2012} and by perturbation theory~\cite{Pastukhov2018},
a truncated-basis variational approach~\cite{Levinsen2019, Amelio2023, Nakano2024},
diffusion quantum Monte-Carlo calculations~\cite{Akaturk2019, Pena_Ardila2020}, functional renormalization group theory~\cite{Isaule2021},
a $T$-matrix approximation~\cite{CardenasCastillo2022},
and variational mean-field (coherent-state) methods, both static and dynamic~\cite{Hryhorchak2020, Panochko2022, Shi2024}.

The problem of a Bose polaron in a dense excitonic system with realistic interaction laws [such as Eq.~\eqref{eqn:V_IX}] has received much less attention.
Some nonperturbative calculations within the hypernetted chain method 
have been reported~\cite{Xu2021}.
Unfortunately, those results are not directly relevant for the present study because of a different geometry of the problem (an e-h quadrilayer instead of the bilayer).

In general, the goal is to find the dispersion $E = E(P)$ of the Bose polarons,
which is determined by the peaks of the spectral function
\begin{equation}
A_\mathrm{DX}(P, E) = -2\, \mathrm{Im}\, G_\mathrm{DX}(P, E)\,,
\label{eqn:A}
\end{equation}
where
\begin{equation}
\begin{split}
	G_\mathrm{DX}(P, E) &=
	-i \int\limits_0^\infty d t e^{i E t / \hbar} \left\langle \left[a_\mathbf{P}^{\phantom\dagger}(t), a_{\mathbf{P}}^{\dagger}(0)
	\right] \right\rangle\\
	&\equiv
	\left[E - \frac{\hbar^2 P^2}{2 m_\mathrm{DX}} - \Sigma(P, E) + i 0^+ \right]^{-1}
	\label{eqn:Green_pole}
\end{split}
\end{equation}
is the retarded Green's function of the impurity (in our case, a DX) and
$a_{\mathbf{k}}^{\phantom\dagger} (a_\mathbf{k}^\dagger)$ is the impurity creation (annihilation) operators.
To analyze the polaron resonances probed in optical experiments it is sufficient to consider $P = 0$ only, and so we suppress the momentum argument $P$ in the formulas below.

Within the $T$-matrix method the self-energy of the Bose polaron is given by
\begin{equation}
\Sigma(E) = n T(E)\,,
\label{eqn:Sigma_from_T}
\end{equation}
which is similar to Eq.~\eqref{eqn:zeta_from_n}.
A formula for the $T$-matrix of a weakly-coupled BEC of spinless bosons
has been proposed by Raith and Schmidt (RS)~\cite{Rath2013}.
In our notations, it looks as follows:
\begin{align}
	T(E) &= \frac{\widetilde{V}}{1 - L_\mathrm{RS}(E) \widetilde{V}}\,,
	\label{eqn:T_Rath}\\
	L_\mathrm{RS}(E) &= 
	\frac{1}{\Omega} \sum_{|\mathbf{k}| < \Lambda}
	\frac{u_\mathbf{k}^2}{E - \omega_\mathbf{k} - \varepsilon_{\mathrm{DX}, \mathbf{k}}}\,,
	\label{eqn:L_Rath}
\end{align}
where
\begin{align}
	\omega_\mathbf{k} &= \sqrt{\varepsilon_\mathbf{k}^2
		+ 2 \zeta \varepsilon_\mathbf{k}^{\phantom\dagger}}\,,
\label{eqn:omega_k}\\
	u_\mathbf{k} &=\frac12 \left(\sqrt{\frac{\omega_\mathbf{k}}{\varepsilon_\mathbf{k}}}
	+ \sqrt{\frac{\varepsilon_\mathbf{k}}{\omega_\mathbf{k}}}\,\,
	\right)
\label{eqn:v}
\end{align}
are the Bogoliubov excitation energies and coherence factors.
RS derived Eq.~\eqref{eqn:T_Rath} by summing a subset of ladder diagrams.
Identical expressions have been also obtained within the truncated-basis approach~\cite{Nakano2024}.
Focusing on the equal-mass case $m_\mathrm{DX} = m$,
it is easy to show analytically that $L_\mathrm{RS}(E) = L(E)$ [cf.~Eq.~\eqref{eqn:L}].
Hence,
these theories predict, surprisingly, that $T(E)$ is no different from the vacuum two-body $T$-matrix given by Eq.~\eqref{eqn:T^u}.
Therefore, to adapt this approach to the spinful case, we can
use our results from Sec.~\ref{sub:DX-IX} and try
\begin{equation}
	T(E) = \frac34\, T^\mathrm{a}(E) + \frac14\, T^\mathrm{s}(E)\,,
	\label{eqn:T_two_spins_NSCT}
\end{equation}
assuming equal concentrations of all IX spin states.

In our model the triplet term $T^\mathrm{a}(E) = \widetilde{V}_x$
is energy-independent, and so it shifts the self-energy by a fixed amount
\begin{equation}
\Delta\Sigma = \frac34\, \widetilde{V}_x n\,,
\label{eqn:Sigma_a}
\end{equation}
which is equivalent to a shift of the DX chemical potential.
This suggests an improved approximation
\begin{align}
\Sigma(E) &= \Delta\Sigma + \frac14\, n T^\mathrm{s}(E - \Delta\Sigma)
	\notag\\
&= \Delta\Sigma + \frac{\pi \hbar^2}{m} n \left[\ln\left(-\frac{E_\mathrm{XX}}{E - \Delta\Sigma} \right)\right]^{-1}.
	\label{eqn:T_two_spins_SCT}
\end{align}
(We used $2\mu = m$ in the denominator assuming $m_\mathrm{DX} = m$.)
The resultant spectral function $A_\mathrm{DX}(E)$ has peaks at energies that solve the equation $E = \mathrm{Re}\, \Sigma(E)$.
The higher-energy solution is the RBP: 
\begin{equation}
	E_\mathrm{RBP} \simeq \Delta\Sigma(n) + \frac{\pi\hbar^2}{m}
	\frac{n}{\ln \left(1 / n a_\mathrm{XX}^2\right)}\,,
	\quad
	n \ll a_\mathrm{XX}^{-2}\,.
	\label{eqn:E_RBP_small_n}
\end{equation}
This equation is different from those previously derived for spinless bosons~\cite{Pastukhov2018, Pena_Ardila2020} in two aspects.
One is the addition of $\Delta\Sigma(n)$, the other is the extra factor of $1/4$ in the second term. Both differences originate from the electron spin.
The `repulsive' nature of the RBP is manifested in its energy increase with $n$,
which is due to the positive sign of $\mathrm{Re}\,T(E)$.
Note that $\mathrm{Re}\,T^\mathrm{s}(E) > 0$
at $-E_{\mathrm{XX}} < E < E_{\mathrm{XX}}$, which can be thought of as a `level repulsion' at energies above the bound-state resonance.
At the face value, Eq.~\eqref{eqn:E_RBP_small_n} predicts a diverging $E_\mathrm{RBP}$ at $n \to 1 / a_\mathrm{XX}^2$.
This is referred to as the strong coupling regime for the Bose polaron.
In fact, at large $n$, this solution of the equation $E = \mathrm{Re}\, \Sigma(E)$ has the asymptotic behavior $E_{\mathrm{RBP}} \simeq \Delta\Sigma(n) + E_{\mathrm{XX}}$.

The lower-energy solution corresponds to the ABP.
It depends on $n$ as
\begin{subequations} \label{eqn:E_ABP}
	\begin{numcases}{E_\mathrm{ABP} \simeq \Delta\Sigma(n) - }
		\frac{\pi\hbar^2}{m} n + E_\mathrm{XX}\,, &  $n \ll a_\mathrm{XX}^{-2}$\,,
		\label{eqn:E_ABP_small_n}\\
		\frac{\pi\hbar^2}{m}
		\frac{n}{\ln \left(n a_\mathrm{XX}^2\right)}\,, & $n \gg a_\mathrm{XX}^{-2}$\,.
		\label{eqn:E_ABP_large_n}
	\end{numcases}
\end{subequations}
Note that Eq.~\eqref{eqn:E_ABP_large_n} is the same as Eq.~\eqref{eqn:E_RBP_small_n}.
However, the `reduced energy' $E_\mathrm{ABP} - \Delta\Sigma(n)$ now decreases with $n$, which is a signature of DX-IX attraction.

\begin{figure}
	\begin{center}
		\includegraphics[width=8cm]{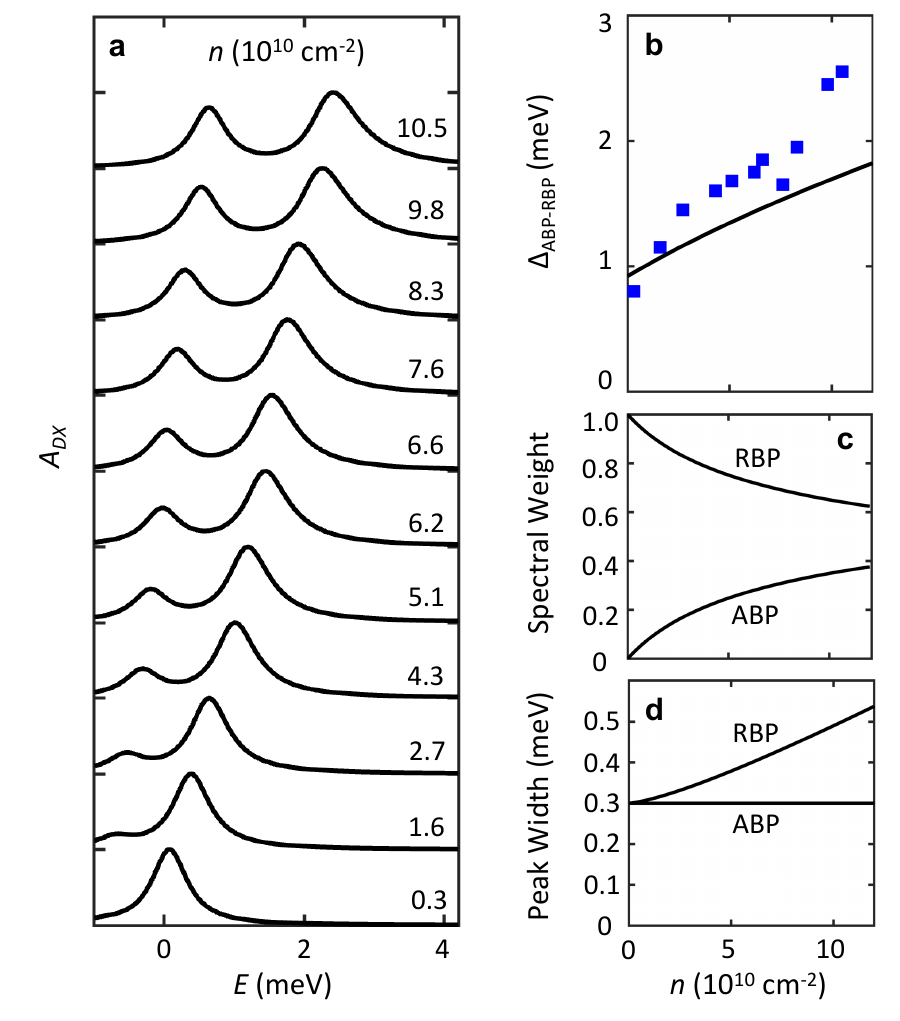}
		\caption{(a) Calculated DX spectral function for different $n$ using $E_\mathrm{XX} = (0.96 + 0.88) / 2 = 0.92\,\mathrm{meV}$ and damping $\Gamma = 0.3\,\mathrm{meV}$. The lower-energy and higher-energy peaks correspond to the ABP and RBP, respectively.
			(b) The ABP-RBP energy splitting deduced from panel (a) (line). The squares are experimental data from Fig.~2 of the main text. 
			(c) ABP and RBP spectral weight vs. $n$. (d) ABP and RBP peak width vs. $n$.
			\label{fig:spectra4}
		}
	\end{center}
\end{figure}

The DX spectral function computed numerically from Eqs.~\eqref{eqn:A}, \eqref{eqn:Green_pole},
\eqref{eqn:Sigma_a}, and \eqref{eqn:T_two_spins_SCT} is plotted in Fig.~\ref{fig:spectra4}(a).
To regularize the $\delta$-function-like ABP peak we added a damping constant $-i \Gamma$ to $\Delta\Sigma$.
Both the ABP and RBP energies increase with IX density $n$, in a qualitative agreement with the experiment. The rate of increase is however somewhat smaller. The distance $\Delta_\mathrm{ABP-RBP}$ between the two peaks as a function of $n$ is shown in
Fig.~\ref{fig:spectra4}(b). The starting point, $\Delta_\mathrm{ABP-RBP} = E_\mathrm{XX}$ is in a good agreement with the measured value, the subsequent rate of increase is about twice slower.
In the context of the polaron problem, the integrated weight (or so-called quasiparticle residue) of the spectral peaks is often discussed.
As shown in Fig.~\ref{fig:spectra4}(c),
the spectral weight is steadily transferred from the RBP to to ABP as $n$ increases,
which is also apparent from Fig.~\ref{fig:spectra4}(a).
Finally, in Fig.~\ref{fig:spectra4}(d) we present the evolution of the peak widths.
The ABP peak maintains the constant width equal to $\Gamma$ (which we added by hand).
The RBP peak widens with $n$.
This widening originates from the imaginary part of the $T$-matrix and represents collisional broadening of an unbound DX being scattered by IXs.

%% FIG. 4

The described $T$-matrix theory is certainly an approximation.
It does not capture several additional effects as follows.
In Sec.~\ref{sub:DX-IX} we suggested that the ABP is essentially a dimer.
In fact, the ABP can still be dressed with Bogoliubov-like excitations of the medium, i.e.,  density oscilations localized near the dimer.
Such excitations would produce spectral weight above the lowest-energy ABP state.
This spectral weight can be substantial. In the strong-coupling polaronic regime,
it may even exceed that of the ground ABP state.
Conversely, for the RBP, which is a metastable state,
these local modes typically have negative energies,
producing spectral lines below the main RBP peak~\cite{Shchadilova2016}.
Therefore, a non-negligible absorption can be present everywhere in between ABP and RBP energies.

%%%%%%%%%%%%%%%%%%%%%%%%%%%%%%%%%%%%%%%%%%%%%%%%%%%%%%%%%%%%%%%%%%%%%%%%%%%%%%%%%%%%%%
%%%%%%%%%%%%%%%%%%%%%%%%%%%%%%%%%%%%%%%%%%%%%%%%%%%%%%%%%%%%%%%%%%%%%%%%%%%%%%%%%%%%%%
%%%%%%%%%%%%%%%%%%%%%%%%%%%%%%%%%%%%%%%%%%%%%%%%%%%%%%%%%%%%%%%%%%%%%%%%%%%%%%%%%%%%%%
\section{A phenomenological $\mathbf{T}$-matrix model}
\label{sec:g-ology}

The $T$-matrix theory of Sec.~\ref{sec:BEC} gives a qualitative but not quantitative agreement with the experiment.
It is also not fully satisfactory for several conceptual reasons.
First, Eq.~\eqref{eqn:L_Rath} disagrees with the perturbation theory formula~\cite{Christensen2015, Pastukhov2018}
\begin{equation}
	\Sigma = n\widetilde{V} + 
	\frac{n\widetilde{V}^2}{\Omega} \sum_{|\mathbf{k}| < \Lambda}
	\frac{\varepsilon_\mathbf{k}}{\omega_\mathbf{k}}\,
	\frac{1}{E - \omega_\mathbf{k} - \varepsilon_{\mathrm{DX}, \mathbf{k}}}
	\label{eqn:Sigma_Frohlich}
\end{equation}
already in the order $O(\widetilde{V}^2)$ unlike
other theoretical calculations~\cite{Shchadilova2016, Hryhorchak2020},
which do agree with Eq.~\eqref{eqn:Sigma_Frohlich}.
The perturbation theory indicates that the response of the BEC
to the impurity is suppressed at energy scales below $\zeta$ where it behaves as a fairly `rigid' medium with excitation energies much larger than the bare particle energies, $\omega_\mathbf{k} \gg \varepsilon_\mathbf{k}$.
In contrast, 
the RS theory~\cite{Rath2013} and the truncated-basis method~\cite{Nakano2024} (at the single-Bogoliubov-excitation level)
predict that the interaction among host bosons practically do not affect the response of the BEC. (If $m_\mathrm{DX} = m$, there is no difference at all, see Sec.~\ref{sec:BEC}.)

Second, as explained in Sec.~\ref{sub:IX-IX}, the IX system is strongly correlated,
so diagrammatic approaches, perturbative or otherwise, are uncontrolled.
In the same vein, formulas like Eqs.~\eqref{eqn:L_Rath} or \eqref{eqn:Sigma_Frohlich}
assume unrealistic (extremely short-range) IX-IX interaction law.

It may therefore be prudent to retain only the basic properties of the
theory outlined in the previous section and
make phenomenological assumptions about all quantities that are difficult to compute reliably.
Returning to Eq.~\eqref{eqn:T_two_spins_SCT},
we can argue that it represents splitting of the self-energy into
a non-singular part with a slow $E$-dependence and a singular part that
has a pole at some energy
\begin{equation}
E_\mathrm{ABP}^{(0)} = -E_\mathrm{XX} + n g_2\,.
\label{eqn:E_ABP_0}
\end{equation}
This leads us to the model introduced in the main text:
\begin{equation}
	T(E) = g_1 + g_3 \frac{E_\mathrm{XX}}{E - E_\mathrm{ABP}^{(0)}} .
	\label{eqn:T-matrix}
\end{equation}
As stated therein, this model predicts the polaron energies 
\begin{align}
	E_\mathrm{ABP,RBP} &= \frac12\, (2 n g_1 - E_\mathrm{XX}
	\pm \Delta_\mathrm{ABP-RBP} )\,,
	\label{eqn:mean-field1}	\\
	\Delta_\mathrm{ABP-RBP} &= \left(E_\mathrm{XX}^2
	+ 4 n g_3 E_\mathrm{XX}
	\right)^{1/2},
	\label{eqn:Delta}
\end{align}
which agree fairly well with the measured peak energies.
Here we already set $g_1 = g_2$ because it is physically reasonable
if the DX-IX biexciton is weakly bound and because
it helps to reduce the number of phenomenological parameters.
This model also predicts the polaron spectral weights (quasiparticle residues)
\begin{equation}
	Z_\mathrm{ABP,RBP} 
	= \frac{1}{1 - (d\Sigma / dE)}
	= \frac12 \pm \frac12\,
	\frac{E_\mathrm{XX}}{\Delta_\mathrm{ABP-RBP}}\,,
	\label{eqn:Z}
\end{equation}
which depend on $n$ similar to what is shown in Fig.~\ref{fig:spectra4}(c).

We can use the formulas of Secs.~\ref{sub:DX-IX} and \ref{sec:BEC} to crudely estimate $g_1$ and $g_3$.
For the case of $g_1$, we take $n g_1 = \Delta\Sigma(n)$, i.e.,
$g_1 = (3/4) \widetilde{V}_x$, see Eq.~\eqref{eqn:Sigma_a}.
For $g_3$, we use Eqs.~\eqref{eqn:T^s} and \eqref{eqn:T_two_spins_NSCT}
to obtain $g_3 = \pi \hbar^2 / 2\mu$.
These are the estimates quoted in the main text, e.g.,
$g_1 = 0.21 \times 10^{-10}\,\mathrm{meV}\, \mathrm{cm}^2$ for hh.
It is also possible to extract $g_1$
from the measured peak positions by fitting them
to Eqs.~\eqref{eqn:mean-field1} and \eqref{eqn:Delta}.
Doing so for the hh points in Fig.~2b, we obtained
$g_1 = 0.34 \times 10^{-10}\,\mathrm{meV}\, \mathrm{cm}^2$.
% We have also tried multi-parameter fitting using all $g_i$'s as adjustable parameters.
% We found however that this procedure is not very reliable as fits of comparable
% good quality can be produced by rather different combinations of parameter values.
A better physical understanding of these parameter values and other spectral characteristics of the excitonic Bose polarons warrants future experimental and theoretical work.

%%%%%%%%%%%%%%%%%%%%%%%%%%%%%%%%%%%%%%%%%%%%%%%%%%%%%%%%%%
%%%%%%%%%%%%%%%%%%%%%%%%%%%%%%%%%%%%%%%%%%%%%%%%%%%%%%%%%%
%%%%%%%%%%%%%%%%%%%%%%%%%%%%%%%%%%%%%%%%%%%%%%%%%%%%%%%%%%
\bibliography{Excitonic_Bose-polarons}